\documentclass[journal,comsoc]{IEEEtran}
\usepackage[T1]{fontenc}

\ifCLASSINFOpdf
   \usepackage[pdftex]{graphicx}
\else
\fi
\usepackage{amsmath}
\usepackage[cmintegrals]{newtxmath}
\newtheorem{theorem}{Theorem}
\newtheorem{definition}{Definition}

\newtheorem{proposition}{Proposition}
\newtheorem{remark}{Remark}
\usepackage{bm}
\usepackage{algorithm}  
\usepackage{algorithmicx}  
\usepackage{algpseudocode}  
\usepackage{amsmath}  

\usepackage{float}  
\usepackage{lipsum}
\makeatletter
\usepackage{array}
\usepackage{multirow}
\usepackage{longtable}
\usepackage{rotating}
\usepackage{stfloats}
\usepackage{graphicx}  
\ifCLASSOPTIONcompsoc
\usepackage[caption=false,font=normalsize,labelfon
t=sf,textfont=sf]{subfig}
\else
\usepackage[caption=false,font=footnotesize]{subfig}
\fi

\begin{document}
%
\title{Tensor Recovery Based on A Novel Non-convex Function Minimax Logarithmic Concave Penalty Function}
%
%
%

\author{Hongbing Zhang,
	Xinyi Liu, Chang Liu, Hongtao Fan, Yajing Li, Xinyun Zhu
	\thanks{This work was supported by the National Natural Science Foundation of
		China (Nos. 11701456, 11801452, 11571004), Fundamental Research Project of Natural Science in Shaanxi Province General Project (Youth) (Nos. 2019JQ-415, 2019JQ-196), the Initial Foundation for Scientific Research of Northwest A\&F University (Nos.2452017219, 2452018017), and Innovation and Entrepreneurship Training Program for College Students of Shaanxi Province (S201910712132).}
	\thanks{H. Zhang, X. Liu, C. Liu, H. Fan and Y. Li are with Department of information and Computing Science, College of Science, Northwest A\&F University, Yangling, Shaanxi 712100, China(e-mail: zhanghb@nwafu.edu.cn; Lxy6x1@163.com; chang\_l0923@163.com
		; fanht17@nwafu.edu.cn; hliyajing@163.com).}
	\thanks{X. Zhu is with the Department of Mathematics, University of Texas of the Permian Basin, Odessa, TX 79762, USA(e-mail: zhu\_x@utpb.edu).
	}
}
%
%

\markboth{}%
{Shell \MakeLowercase{\textit{et al.}}: Bare Demo of IEEEtran.cls for IEEE Communications Society Journals}
%



\maketitle

\begin{abstract}
Non-convex relaxation methods have been widely used in tensor recovery problems, and compared with convex relaxation methods,  can achieve better recovery results. In this paper, a new non-convex function, Minimax Logarithmic Concave Penalty (MLCP) function, is proposed, and some of its intrinsic properties are analyzed, among which it is interesting to find that the Logarithmic function is an upper bound of the MLCP function. The proposed function is generalized to tensor cases, yielding tensor MLCP and weighted tensor $L\gamma$-norm. Consider that its explicit solution cannot be obtained when applying it directly to the tensor recovery problem. Therefore, the corresponding equivalence theorems to solve such problem are given, namely, tensor equivalent MLCP theorem and equivalent weighted tensor $L\gamma$-norm theorem. In addition, we propose two EMLCP-based models for classic tensor recovery problems, namely low-rank tensor completion (LRTC) and tensor robust principal component analysis (TRPCA), and design proximal alternate linearization minimization (PALM) algorithms to solve them individually. Furthermore, based on the Kurdyka-Łojasiwicz property, it is proved that the solution sequence of the proposed algorithm has finite length and converges to the critical point globally. Finally, Extensive experiments show that proposed algorithm achieve good results, and it is confirmed that the MLCP function is indeed better than the Logarithmic function in the minimization problem, which is consistent with the analysis of theoretical properties.
\end{abstract}

\IEEEpeerreviewmaketitle

\section{Introduction}

\IEEEPARstart{D}{ata} structures become more complex, and the processing required by many applications becomes more difficult as the dimensionality of the data increases. As a representation of multi-dimensional data, tensors have played an important role in many high-dimensional data applications in recent years, such as color image/video (CI/CV) processing \cite{1472018163}, \cite{3762018t397}, \cite{7562018767}, \cite{1372020149}, hyperspectral/multispectral image (HSI/MSI) processing \cite{8359412}, \cite{7467446}, \cite{8657368}, \cite{8941238}, magnetic resonance imaging (MRI) data recovery \cite{341202131}, \cite{1242020783}, \cite{4032018417}, \cite{9412019964}, video foreground and background subtraction\cite{5032019109}, \cite{8454775}, \cite{8319458}, \cite{7488247}, video rain stripe removal \cite{8237537}, \cite{8578793}, and signal reconstruction \cite{7010937}, \cite{7676397}.

These practical application problems above can be transformed into tensor recovery problems. For different observation data, the tensor recovery problem can usually be modeled as a low-rank tensor completion (LRTC) problem and a tensor robust principal component analysis (TRPCA) problem. Their corresponding models are as follows:
\begin{eqnarray}
\min_{\mathcal{Z}} rank(\mathcal{Z})\quad s.t.P_{\Omega}(\mathcal{T})=P_{\Omega}(\mathcal{Z})\label{LRTC}
\\\min_{\mathcal{Z},\mathcal{E}} rank(\mathcal{Z})+\tau_{1}\|\mathcal{E}\|_{1}\quad s.t. \mathcal{T}=\mathcal{Z}+\mathcal{E},\label{TRPCA}
\end{eqnarray}

where $\mathcal{T}\in\mathbb{R}^{\mathit{I}_{1}\times\mathit{I}_{2}\times\mathit{I}_{3}}$ is the obesrvation; $\mathcal{Z}$ is initial tensor; $\mathcal{E}$ is sparsity tensor; $\mathcal{P}_{\Omega}(\mathcal{Z})$ is a projection
operator that keeps the entries of $\mathcal{Z}$ in $\Omega$ and sets all others to zero. Let
\begin{eqnarray}
\Phi_{\mathbb{G}}(\mathcal{Z}):=\left\{\begin{array}{l}
0,\qquad if\quad\mathcal{Z}\in\mathbb{G},
\\\infty,\qquad otherwise
\end{array}
\right.
\end{eqnarray}
where $\mathbb{G}:=\{\mathcal{Z}\in\mathbb{R}^{\mathit{I}_{1}\times\mathit{I}_{2}\times\mathit{I}_{3}},\mathcal{P}_{\Omega}(\mathcal{Z}-\mathcal{T})=0\}$.

It is not difficult to find that (\ref{LRTC}) and (\ref{TRPCA}) are the problem of solving tensor rank minimization. As we all know, the most popular tensor recovery method is nuclear norm minimization. However, the definition of the rank of a tensor is not unique, different tensor rank and corresponding nuclear norm can be induced based on different tensor decomposition. The CANDECOMP/PARAFAC (CP) rank is equal to the smallest number of rank-1 tensors to achieve CP decomposition \cite{kiers2000towards}, but generally NP-hard to estimate accurately \cite{hillar2013most}. Another popular rank is the Tucker rank derived from the Tucker decomposition \cite{tucker1966some}, which is defined as a vector whose $i$th element corresponds to the rank of the mode $i$ unfolding matrix of tensor. Liu et al. \cite{liu2012tensor} first proposed sum of nuclear norms (SNN) as a convex surrogate of Tucker rank, which significantly facilitated the development of the tensor recovery problem. But the SNN is not compact convex relaxation of Tucker rank, and this matrixing technique cannot fully exploit tensor structure information \cite{romera2013new}. Furthermore, tensor tubal rank and multi-rank are obtained from tensor singular value decomposition (t-SVD) \cite{6416568}. Since there is no need tensor matrixization in the calculation process, this allows better utilization of the tensor's internal structural information. Many multidimensional data in the real world can be well approximated by low-rank tensors, due to the fact that the singular values of the corresponding tensors are relatively small, while a few large ones contain the main information. On this basis, the tensor nuclear norm (TNN) has been proposed as a convex relaxation of tubal-rank \cite{7782758}. Recently, Zheng et al. \cite{2020170} proposed a new form of rank (N-tubal rank) based on tubal rank, which adopts a new unfold method of higher-order tensors into third-order tensors in various directions. Benefiting from this, t-SVD can be applied to higher-order situations by solving third-order tensors of various directions forms. This approach not only enables t-SVD to be applied to higher order cases, but also makes good use of the properties of tensor tubal rank. In view of the excellent properties of N-tubal rank, we will use N-tubal rank to construct the model in this paper.

However, although nuclear norm relaxation is becoming a popular solution to the low rank tensor recovery problem, it still suffers from some drawbacks. TNN is a convex relaxation approximation of tensor tubal rank, and there is still a certain distance from tensor tubal rank minimization, which usually leads to the solution of the optimization problem being suboptimal solution to the original problem. Recently, to break the limitation of biased estimation of convex relaxation methods, some non-convex relaxation strategies have been adopted to solve the tensor recovery problem. Non-convex methods are able to penalize larger singular values less and smaller singular values more. In \cite{kong2018t}, a t-Schatten-p norm was proposed to approximate tensor tubal rank by extending the Schatten-p norm. Another non-convex approach to approximating the tensor tubal rank is by transforming each element as a sum of t-TNNs with the Laplace function \cite{xu2019laplace}. Besides, Logarithmic function \cite{gong2013general}, MCP function \cite{you2019nonconvex}, \cite{zhang2010nearly}, \cite{2132021245}, SCAD function \cite{fan2001variable} are also applied to carry out non-convex relaxation. To further explore the superiority of non-convex functions and improve the accuracy of tensor recovery, we propose a new non-convex function in this paper, namely the Minimax Logarithmic Concave Penalty (MLCP) function. Interestingly, it is found that the Logarithmic function is an upper bound of the MLCP function. The proposed function is then generalized to higher dimensional cases, yielding vector MLCP, matrix MLCP, tensor MLCP, and weighted tensor $L\gamma$-norm. However, when the proposed function is directly applied to the tensor recovery problem, the explicit solution cannot be obtained, which is very unfavorable to the solution of the algorithm. Therefore, we further put forward the corresponding equivalence theorems, namely vector equivalent MLCP theorem, matrix equivalent MLCP theorem, tensor equivalent MLCP theorem, and equivalent weighted tensor $L\gamma$-norm theorem, to tackle this problem. Furthermore, we give the proximal operator for the equivalent weighted tensor $L\gamma$-norm, which makes the tensor recovery model easier to solve. Finally, similar to the technique employed in \cite{qiu2021nonlocal}, \cite{zhang2018nonconvex}, \cite{zhang2017nonconvex}, \cite{ochs2015iteratively}, using the Proximal Alternating Linearization Minimization Algorithm (PALM) \cite{bolte2014proximal}, combined with the Kurdyka-Łojasiwicz property, we prove that the solution sequence obtained with MLCP functions has a finite length and converges globally to the critical point with stronger convergence. 

In summary, the main contributions of our paper are:

First, a new non-convex function, Minimax Logarithmic Concave Penalty (MLCP) function, is proposed. It is found that the Logarithmic function is an upper bound of the MLCP function. The function is generalized to tensor cases, yielding tensor MLCP and weighted tensor $L\gamma$-norm. Considering that applying it directly to the tensor recovery problem, the explicit solution cannot be obtained, which is very unfavorable for the solution of the algorithm. For this reason, we give the corresponding equivalence theorems to solve this problem, namely tensor EMLCP, and equivalent weighted tensor $L\gamma$-norm theorems. The properties of the tensor EMLCP and the equivalent weighted tensor $L\gamma$-norm are analyzed. Furthermore, the proximal operator for the equivalent weighted tensor $L\gamma$-norm is given, so as to make the tensor recovery model easier to solve.

Second, we construct corresponding EMLCP-based models for two typical problems of tensor recovery, and design a Proximal Alternating Linearization Minimization Algorithm (PALM) to solve these two EMLCP-based models. In particular, we adopt a model that removes mixed noise for the TRPCA problem, which is more realistic. Furthermore, based on the Kurdyka-Łojasiwicz property, it is proved that the solution sequence of the proposed algorithm has finite length and converges to the critical point globally.

Third, we conduct experiments on both LRTC and TRPCA using real data. The LRTC experiments on HSI, MRI, CV and the TRPCA experiments on HSI demonstrate the effectiveness of our proposed new non-convex relaxation method. This method yields better results than the Logarithmic relaxation method, which is consistent with our theoretical analysis.

The summary of this article is as follows: In Section II, some preliminary knowledge and background are given. The definitions and theorems of the MLCP function are presented in Section III. In Section IV, we establish the EMLCP-based models and algorithms. In Section V, we give the theoretical convergence analysis of the proposed algorithms. The results of extensive experiments and discussion are presented in Section VI. Conclusions are drawn in Section VII.

\section{Prelimiaries}
\subsection{Tensor Notations and Definitions}
In this section, we give some basic notations and briefly introduce some definitions used throughout the paper. Generally, a lowercase letter and an uppercase letter denote a vector $z$ and a marix $Z$, respectively. An $N$th-order tensor is denoted by a calligraphic upper case letter $\mathcal{Z}\in \mathbb{R}^{\mathit{I}_{1}\times\mathit{I}_{2}\times\cdots\times\mathit{I}_{N}}$ and $\mathit{z}_{i_{1},i_{2},\cdots,i_{N}}$ is its $(i_{1},i_{2},\cdots,i_{N})$-th element. The Frobenius norm of a tensor is defined as $\|\mathcal{Z}\|_{F}=(\sum_{i_{1},i_{2},\cdots,i_{N}}\mathit{y}_{i_{1},i_{2},\cdots,i_{N}}^{2})^{1/2}$. For a three order tensor $\mathcal{Z}\in\mathbb{R}^{\mathit{I}_{1}\times\mathit{I}_{2}\times\mathit{I}_{3}}$, we use $\bar{\mathcal{Z}}$ to denote the discrete Fourier transformation (DFT) along each tubal of $\mathcal{Z}$, i.e., $\bar{\mathcal{Z}}=fft(\mathcal{Z},[],3)$. The inverse DFT is computed by command $ifft$ satisfying $\mathcal{Z}=ifft(\bar{\mathcal{Z}},[],3)$. More often, the frontal slice $\mathcal{Z}(:,:,i)$ is denoted compactly as $\mathcal{Z}^{(i)}$.
\begin{definition}[Mode-$k_{1}k_{2}$ slices \cite{2020170}]
	For an $N$th-order tensor $\mathcal{Z}\in \mathbb{R}^{\mathit{I}_{1}\times\mathit{I}_{2}\times\cdots\times\mathit{I}_{N}}$, its mode-$k_{1}k_{2}$ slices ($\mathcal{Z}^{(k_{1}k_{2})},1\leqslant k_{1} <k_{2}\leqslant N,k_{1},k_{2}\in\mathbb{Z}$) are two-dimensional sections, defined by fixing all but the mode-$k_{1}$ and the mode-$k_{2}$ indexes.
\end{definition}
\begin{definition}[Tensor Mode-$k_{1},k_{2}$ Unfolding and Folding \cite{2020170}]
	For an $N$th-order tensor $\mathcal{Z}\in \mathbb{R}^{\mathit{I}_{1}\times\mathit{I}_{2}\times\cdots\times\mathit{I}_{N}}$, its mode-$k_{1}k_{2}$ unfolding is a three order tensor denoted by $\mathcal{Z}_{(k_{1}k_{2})}\in\mathbb{R}^{\mathit{I}_{k_{1}}\times\mathit{I}_{k_{2}}\times\prod_{s\neq k_{1},k_{2}}\mathit{I}_{s}}$, the frontal slices of which are the lexicographic orderings of the mode-$k_{1}k_{2}$ slices of $\mathcal{Z}$. Mathematically, the  $(i_{1}, i_{2},...,i_{N} )$-th element of $\mathcal{Z}$ maps to the $(i_{k_{1}},i_{k_{2}},j)$-th element of $\mathcal{Z}_{(k_{1}k_{2})}$, where
	\begin{equation}
	j=1+\sum_{s=1,s\neq k_{1},s\neq k_{2}}^{N}(i_{s}-1)\mathit{J}_{s}\quad with\quad \mathit{J}_{s}=\prod_{m=1,m\neq k_{1},m\neq k_{2}}^{s-1}\mathit{I}_{m}. 
	\end{equation}
	The mode-$k_{1}k_{2}$ unfolding operator and its inverse operation are respectively denoted as $\mathcal{Z}_{(k_{1}k_{2})}:=t-unfold(\mathcal{Z},k_{1},k_{2})$ and $\mathcal{Z}:=t-fold(\mathcal{Z}_{(k_{1}k_{2})},k_{1},k_{2})$.
\end{definition}

For a three order tensor $\mathcal{Z}\in\mathbb{R}^{\mathit{I}_{1}\times\mathit{I}_{2}\times\mathit{I}_{3}}$, the block circulation operation is defined as
\begin{equation}
bcirc(\mathcal{Z}):=
\begin{pmatrix}
Z^{(1)}& Z^{(\mathit{I}_{3})}&\dots& Z^{(2)}&\\
Z^{(2)}& Z^{(1)}&\dots& Z^{(3)}&\\
\vdots&\vdots&\ddots&\vdots&\\
Z^{(\mathit{I}_{3})}& Z^{(\mathit{I}_{3}-1)}&\dots& Z^{(1)}&
\end{pmatrix}\in\mathbb{R}^{\mathit{I}_{1}\mathit{I}_{3}\times\mathit{I}_{2}\mathit{I}_{3}}.\nonumber
\end{equation}

The block diagonalization operation and its inverse operation are defined as 
\begin{eqnarray}
&&bdiag(\mathcal{Z}):=\begin{pmatrix}
Z^{(1)} & & &\\
& Z^{(2)} & &\\
& & \ddots &\\
& & & Z^{(\mathit{I}_{3})}
\end{pmatrix} \in\mathbb{R}^{\mathit{I}_{1}\mathit{I}_{3}\times\mathit{I}_{2}\mathit{I}_{3}},\nonumber\\
&&bdfold(bdiag(\mathcal{Z})):=\mathcal{Z}.\nonumber
\end{eqnarray}

The block vectorization operation and its inverse operation are defined as 
\begin{eqnarray}
bvec(\mathcal{Z}):=\begin{pmatrix}
Z^{(1)}\\Z^{(2)}\\\vdots\\Z^{(\mathit{I}_{3})}
\end{pmatrix}\in\mathbb{R}^{\mathit{I}_{1}\mathit{I}_{3}\times\mathit{I}_{2}},\quad bvfold(bvec(\mathcal{Z})):=\mathcal{Z}.\nonumber
\end{eqnarray}
\begin{definition}[t-product \cite{6416568}]
	Let $\mathcal{A}\in\mathbb{R}^{\mathit{I}_{1}\times\mathit{I}_{2}\times\mathit{I}_{3}}$ and $\mathcal{B}\in\mathbb{R}^{\mathit{I}_{2}\times\mathit{J}\times\mathit{I}_{3}}$. Then the t-product $\mathcal{A}\ast\mathcal{B}$ is defined to be a tensor of size $\mathit{I}_{1}\times\mathit{J}\times\mathit{I}_{3}$,
	\begin{eqnarray}
	\mathcal{A}\ast\mathcal{B}:=bvfold(bcirc(\mathcal{A})bvec(\mathcal{B})).\nonumber
	\end{eqnarray}
	
	Since that circular convolution in the spatial domain is equivalent to multiplication in the Fourier domain, the t-product between two tensors $\mathcal{C}=\mathcal{A}\ast\mathcal{B}$ is equivalent to
	\begin{eqnarray}
	\bar{\mathcal{C}}=bdfold(bdiag(\bar{\mathcal{A}})bdiag(\bar{\mathcal{B}})).\nonumber
	\end{eqnarray}
\end{definition}
\begin{definition}[Tensor conjugate transpose \cite{6416568}]
	The conjugate transpose of a tensor $\mathcal{A}\in\mathbb{C}^{\mathit{I}_{1}\times\mathit{I}_{2}\times\mathit{I}_{3}}$ is the tensor $\mathcal{A}^{H}\in\mathbb{C}^{\mathit{I}_{2}\times\mathit{I}_{1}\times\mathit{I}_{3}}$ obtained by conjugate transposing each of the frontal slices and then reversing the order of transposed frontal slices 2 through $\mathit{I}_{3}$.
\end{definition}
\begin{definition}[identity tensor \cite{6416568}]
	The identity tensor $\mathcal{I}\in\mathbb{R}^{\mathit{I}_{1}\times\mathit{I}_{1}\times\mathit{I}_{3}}$ is the tensor whose first frontal slice is the $\mathit{I}_{1}\times\mathit{I}_{1}$ identity matrix, and whose other frontal slices are all zeros.
\end{definition}

It is clear that $bcirc(\mathcal{I})$ is the  $\mathit{I}_{1}\mathit{I}_{3}\times\mathit{I}_{1}\mathit{I}_{3}$ identity matrix. So it is easy to get $\mathcal{A}\ast\mathcal{I}=\mathcal{A}$ and $\mathcal{I}\ast\mathcal{A}=\mathcal{A}$. 
\begin{definition}[orthogonal tensor \cite{6416568}]
	A tensor $\mathcal{Q}\in\mathbb{R}^{\mathit{I}_{1}\times\mathit{I}_{1}\times\mathit{I}_{3}}$ is orthogonal if it satisfies
	\begin{eqnarray}
	\mathcal{Q}\ast\mathcal{Q}^{H}=\mathcal{Q}^{H}\ast\mathcal{Q}=\mathcal{I}.\nonumber
	\end{eqnarray}
\end{definition}
\begin{definition}[f-diagonal Tensor \cite{6416568}]
	A tensor is called f-diagonal if each of its frontal slices is a diagonal matrix.
\end{definition}
\begin{theorem}[t-SVD \cite{8606166}]
	Let $\mathcal{Z}\in\mathbb{R}^{\mathit{I}_{1}\times\mathit{I}_{2}\times\mathit{I}_{3}}$ be a three order tensor, then it can be factored as 
	\begin{eqnarray}
	\mathcal{Z}=\mathcal{U}\ast\mathcal{S}\ast\mathcal{V}^{H},\nonumber
	\end{eqnarray}
	where $\mathcal{U}\in\mathbb{R}^{\mathit{I}_{1}\times\mathit{I}_{1}\times\mathit{I}_{3}}$ and $\mathcal{V}\in\mathbb{R}^{\mathit{I}_{2}\times\mathit{I}_{2}\times\mathit{I}_{3}}$ are orthogonal tensors, and $\mathcal{S}\in\mathbb{R}^{\mathit{I}_{1}\times\mathit{I}_{2}\times\mathit{I}_{3}}$ is an f-diagonal tensor.
\end{theorem}

\begin{definition}[tensor tubal-rank and multi-rank \cite{6909886}]
	The tubal-rank of a tensor $\mathcal{Z}\in\mathbb{R}^{\mathit{I}_{1}\times\mathit{I}_{2}\times\mathit{I}_{3}}$, denoted as $rank_{t}(\mathcal{Z})$, is defined to be the number of non-zero singular tubes of $\mathcal{S}$, where $\mathcal{S}$ comes from the t-SVD of $\mathcal{Z}:\mathcal{Z}=\mathcal{U}\ast\mathcal{S}\ast\mathcal{V}^{H}$. That is 
	\begin{eqnarray}
	rank_{t}(\mathcal{Z})=\#\{i:\mathcal{S}(i,:,:)\neq0\}.
	\end{eqnarray}
	The tensor multi-rank of $\mathcal{Z}\in\mathbb{R}^{\mathit{I}_{1}\times\mathit{I}_{2}\times\mathit{I}_{3}}$ is a vector, denoted as $rank_{r}(\mathcal{Z})\in\mathbb{R}^{\mathit{I}_{3}}$, with the $i$-th element equals to the rank of $i$-th frontal slice of $\mathcal{Z}$.
\end{definition}

\begin{definition}[tensor nuclear norm (TNN)]
	The tensor nuclear norm of a tensor $\mathcal{Z}\in\mathbb{R}^{\mathit{I}_{1}\times\mathit{I}_{2}\times\mathit{I}_{3}}$, denoted as $\|\mathcal{Z}\|_{TNN}$, is defined as the sum of the singular values of all the frontal slices of $\bar{\mathcal{Z}}$, i.e.,
	\begin{eqnarray}
	\|\mathcal{Z}\|_{TNN}:=\sum_{i=1}^{\mathit{I}_{3}}\|\bar{\mathcal{Z}}^{(i)}\|_{\ast}
	\end{eqnarray}
	where $\bar{\mathcal{Z}}^{(i)}$ is the $i$-th frontal slice of $\bar{\mathcal{Z}}$, with $\bar{\mathcal{Z}}=fft(\mathcal{Z},[],3)$.
\end{definition}

\begin{definition}[N-tubal rank \cite{2020170}]
	N-tubal rank of an Nth-order tensor $\mathcal{Y}\in \mathbb{R}^{\mathit{I}_{1}\times\mathit{I}_{2}\times\cdots\times\mathit{I}_{N}}$ is defined as a vector, the elements of which
	contain the tubal rank of all mode-$k_{1}k_{2}$ unfolding tensors, i.e.,
	\begin{eqnarray}
	N-rank_{t}(\mathcal{Y}):=(rank_{t}(\mathcal{Y}_{(12)}),rank_{t}(\mathcal{Y}_{(13)}),\cdots,\nonumber\\rank_{t}(\mathcal{Y}_{(1N)}),rank_{t}(\mathcal{Y}_{(23)}),\cdots,\nonumber rank_{t}(\mathcal{Y}_{(2N)}),\cdots,\\rank_{t}(\mathcal{Y}_{(N-1N)}))\in\mathbb{R}^{N(N-1)/2}.
	\end{eqnarray}
\end{definition}

Next, we will introduce some knowledge related to convergence analysis.
\begin{definition}[Proper function \cite{rockafellar2009variational}]
	Let $\Im$ be a finite-dimensional Euclidean space, a fuction $f:\Im\to[-\infty,+\infty]$ is called proper if $f(z)<+\infty$ for at least one $z\in\Im$, and $f(z)>-\infty$ for all $z\in\Im$.
\end{definition}

The effective domain of $f$ is defined as $dom(f):=\{z:f(z)<+\infty\}$. For a given proper and lower semicontinuous function $f:\Im\to\left( -\infty,+\infty \right] $, the priximal mapping associated with $f$ at $y$ is defined by
\begin{eqnarray}
Prox_{f}(y)=\arg\min_{z\in\Im}\{f(z)+\frac{1}{2}\|z-y\|^{2}\},\quad\forall y\in\Im. \nonumber
\end{eqnarray}
\begin{definition}[Subdifferential of a nonconvex function \cite{rockafellar2009variational}]
	The subdifferential of $f:\mathbb{R}^{n}\to \left( -\infty,+\infty \right]$ at $z$, denoted as $\partial f(z)$, is defined by
	\begin{eqnarray}
	&&\partial f(z)=\{y\in\mathbb{R}^{n}:\quad\exists z^{k}\to z,f(z^{k})\to f(z),\nonumber
	\\&&y^{k}\to y\quad with\quad y^{k}\in\hat{\partial}f(z^{k})\quad as\quad k\to+\infty\},\nonumber
	\end{eqnarray}
	where $\hat{\partial}f(z)$ denotes the Fr\'{e}chet subdifferenetial of $f$ at $z\in dom(f)$, which is the set of all $y$ satisfying
	\begin{eqnarray}
	\lim\inf_{x\to z,x\neq z}\dfrac{f(x)-f(z)-<y,x-z>}{\|x-z\|}\geqslant0.
	\end{eqnarray}
\end{definition}
For any $\mathcal{Z}\in\Re$, the distance from $\mathcal{Z}$ to $\Im$ is defined by $dist(\mathcal{Z},\Im):=\inf\{\|\mathcal{Z}-\mathcal{Y}\|_{F}:\mathcal{Y}\in\Im\}$, where $\Im$ is a subset of $\Re$. Next, we recall the Kurdyka–Łojasiewicz (KL) property, which plays a pivotal role in the analysis of the convergence of proximal alternating linearized minimization (PALM) algorithm for the nonconvex problems.
\begin{definition}[KL function \cite{bolte2014proximal}]
	Let $f:\mathbb{R}^{n}\to \left( -\infty,+\infty \right]$ be a proper and lower semicontinuous function.
	\\\textbf{(a)}:  The function $f$ is said to have the KL property at $z\in dom(\partial f)$ if there exist $\eta\in\left( 0,+\infty \right]$, a	neighborhood $\Im$ of $z$ and a continuous concave function $\phi:\left[ 0,\eta \right) \to \left[ 0,+\infty \right) $ such that: (a) $\phi(0)=0$; (b) $\phi$ is continuously differentiable on $(0,\eta)$, and continuous at $0$; (c) $\phi'(s)>0$ for all $s\in(0,\eta)$; (d) for all $y\in\Im\cap[y\in\mathbb{R}^{n}:f(z)<f(y)<f(z)+\eta]$, the following KL inequality holds:
	\begin{eqnarray}
	\phi'(f(y)-f(z))dist(0,\partial f(y))\geqslant1.\nonumber
	\end{eqnarray}
	\\\textbf{(b)}: If $f$ satisfies the KL property at each point of $dom(\partial f)$, then $f$ is called a KL function.
\end{definition}
\section{Minimax Logarithmic Concave Penalty (MLCP) Function AND EQUIVALENT Minimax Logarithmic Concave Penalty (EMLCP)}
In this section, we first define the definition of the Minimax Logarithmic Concave Penalty (MLCP) function.
\begin{definition}[Minimax Logarithmic Concave Penalty (MLCP) function]
	Let $\lambda>0,\gamma>0,\varepsilon>0$. The MLCP function $f_{L,\gamma,\lambda}:\mathbb{R}\to\mathbb{R}_{\geqslant0}$ is defined as 
	\begin{eqnarray}
	f_{L,\gamma,\lambda}(z)=\left\{\begin{array}{l}
	\lambda \log(\frac{\rvert z\rvert}{\varepsilon}+1)-\frac{\log^{2}(\frac{\rvert z\rvert}{\varepsilon}+1)}{2\gamma}, \rvert z\rvert\leqslant \varepsilon e^{\gamma\lambda}-\varepsilon,
	\\\frac{\gamma\lambda^{2}}{2},\qquad\quad \rvert z\rvert> \varepsilon e^{\gamma\lambda}-\varepsilon.
	\end{array}
	\right.\label{scalarMLCP}
	\end{eqnarray}
\end{definition}

The MLCP function is a symmetric function, so we only discuss its functional properties on $\left[ 0,+\infty \right) $.
\begin{proposition}
	The MLCP function defined in (\ref{scalarMLCP}) satisfies the following properties:
	\\\textbf{(a)}: $f_{L,\gamma,\lambda}(z)$  is continuous, smooth and 
	\begin{eqnarray}
	f_{L,\gamma,\lambda}(0)=0,\lim\limits_{z\to+\infty}\frac{f_{L,\gamma,\lambda}(z)}{z}=0;\nonumber
	\end{eqnarray}
	\textbf{(b)}: $f_{L,\gamma,\lambda}(z)$ is monotonically non-decreasing and concave on $ \left[ 0,+\infty \right) $;
	\\\textbf{(c)}: $f'_{L,\gamma,\lambda}(z)$ is non-negativity and monotonicity non-increasing on $ \left[ 0,+\infty \right) $. Moreover, it is Lipschitz bounded, i.e., there exists constant $L(f)$ such that
	\begin{eqnarray}
	\rvert f'_{L,\gamma,\lambda}(x)-f'_{L,\gamma,\lambda}(y)\rvert\leq L(\ell)\rvert x-y\rvert;\nonumber
	\end{eqnarray}
	\textbf{(d)}: Especially, for the MLCP function, it is increasing in parameter $\gamma$, and
	\begin{eqnarray}
	\lim\limits_{\gamma\to+\infty}f_{L,\gamma,\lambda}(z)=\lambda \log(\frac{\rvert z\rvert}{\varepsilon}+1).
	\end{eqnarray}
\end{proposition}
\begin{IEEEproof}
	The proof is provided in Appendix A.
\end{IEEEproof}
\begin{definition}[Vector MLCP]
	Let $z\in\mathbb{R}^{n}$ and $\lambda>0,\gamma>0,\varepsilon>0$. The vector MLCP $f_{L,\gamma,\lambda}:\mathbb{R}^{n}\to\mathbb{R}_{\geqslant0}$ is defined as 
	\begin{eqnarray}
	f_{L,\gamma,\lambda}(z)=\sum_{i=1}^{n}f_{L,\gamma,\lambda}(z_{i}),
	\end{eqnarray}
	where $z_{i}$ denotes the $i$th entry of the vector $z$ and $f_{L,\gamma,\lambda}(z_{i})$ is defined in (\ref{scalarMLCP}).
\end{definition}
\begin{definition}[Matrix MLCP]
	Let $Z\in\mathbb{R}^{m\times n}$ and $\lambda>0,\gamma>0,\varepsilon>0$. The matrix MLCP $f_{L,\gamma,\lambda}:\mathbb{R}^{m\times n}\to\mathbb{R}_{\geqslant0}$ is defined as
	\begin{eqnarray}
	f_{L,\gamma,\lambda}(Z)=\sum_{i=1}^{m}\sum_{j=1}^{n}f_{L,\gamma,\lambda}(Z_{ij}),
	\end{eqnarray}
	where $Z_{ij}$ denotes the $(i,j)$ element of $Z$, and $f_{L,\gamma,\lambda}$ is the same as in (\ref{scalarMLCP}). 
\end{definition}
\begin{definition}[Tensor MLCP]
	Let $\mathcal{Z}\in\mathbb{R}^{\mathit{I}_{1}\times\mathit{I}_{2}\times\cdots\times\mathit{I}_{N}}$ and $\bar{\lambda}>0,\gamma>0,\varepsilon>0,\bar{\lambda}\in\mathbb{R}^{\mathit{I}_{1}\times\mathit{I}_{2}\times\cdots\times\mathit{I}_{N}}$. The tensor MLCP $f_{L,\gamma,\bar{\lambda}}:\mathbb{R}^{\mathit{I}_{1}\times\mathit{I}_{2}\times\cdots\times\mathit{I}_{N}}\to\mathbb{R}_{\geqslant0}$ is defined as
	\begin{eqnarray}
	f_{L,\gamma,\bar{\lambda}}(\mathcal{Z})=\sum_{i_{1}=1}^{\mathit{I}_{1}}\sum_{i_{2}=1}^{\mathit{I}_{2}}\cdots\sum_{i_{N}=1}^{\mathit{I}_{N}}f_{L,\gamma,\bar{\lambda}_{i_{1},i_{2},\cdots,i_{N}}}(\mathcal{Z}_{i_{1},i_{2},\cdots,i_{N}}),\label{tensormlcp1}
	\end{eqnarray}
	where $\mathcal{Y}_{i_{1},i_{2},\cdots,i_{N}}$ denotes the $(i_{1},i_{2},\cdots,i_{N})$-th element of $\mathcal{Y}$, and $h_{\gamma,\bar{\lambda}}$ is defined in (\ref{scalarMLCP}). 
\end{definition}
\begin{definition}[Matrix $L\gamma$-norm]
	The $L\gamma$ norm of a rank-$r$ matrix $Z\in\mathbb{R}^{m\times n}$, denoted by $\|Z\|_{L,\gamma,\lambda}$, is defined in terms of the singular values \{$\sigma_{i}, i=1,2,\dots,r$\} as follows:
	\begin{eqnarray}
	\|Z\|_{L,\gamma,\lambda}:=f_{L,\gamma,\lambda}(\sigma)=\sum_{i=1}^{r}f_{L,\gamma,\lambda}(\sigma_{i}),
	\end{eqnarray}
	where $\sigma$ is singular value vector of matrix $Z$.
\end{definition}

Similarly, the weighted matrix $L\gamma$-norm is a generalization of weighted MLCP for matrix and is defined as follows.
\begin{definition}[Weighted matrix $L\gamma$-norm]
	The weighted matrix $L\gamma$-norm of $Z\in\mathbb{R}^{m\times n}$, denoted by $\|Z\|_{L,\gamma,\lambda}$, is defined as follows:
	\begin{eqnarray}
	\|Z\|_{L,\gamma,\lambda}=f_{L,\gamma,\lambda}(\sigma)=\sum_{i=1}^{r}f_{L,\gamma,\lambda_{i}}(\sigma_{i}).
	\end{eqnarray}
	where $r=\min(m,n)$ denotes the maximum rank of $Z$.
\end{definition}

\begin{definition}[Weighted tensor $L\gamma$-norm]
	The weighted tensor $L\gamma$-norm of $\mathcal{Z}\in\mathbb{R}^{\mathit{I}_{1}\times\mathit{I}_{2}\times\mathit{I}_{3}}$, denoted by $\|\mathcal{Z}\|_{L,\gamma,\bar{\lambda}}$, is defined as follows:
	\begin{eqnarray}
	\|\mathcal{Z}\|_{L,\gamma,\bar{\lambda}}=\sum_{i=1}^{\mathit{I}_{3}}\|\bar{\mathcal{Z}}^{(i)}\|_{L,\gamma,\bar{\lambda}_{i}}=\sum_{i=1}^{\mathit{I}_{3}}\sum_{j=1}^{R}f_{L,\gamma,\bar{\lambda}_{i,j}}(\sigma_{j}(\bar{\mathcal{Z}}^{(i)})).\label{wtgnmlcp}
	\end{eqnarray}
	where $R=\min(\mathit{I}_{1},\mathit{I}_{2})$.
\end{definition}

Further, we convert $\lambda$ from a constant to a variable, for which we propose some equivalent MLCP theorems.
\begin{theorem}[Scalar EMLCP]
	Let $\lambda>0,\gamma>0,\varepsilon>0$ and $z\in\mathbb{R}$. The scalar MLCP $f_{L,\gamma,\lambda}:\mathbb{R}\to\mathbb{R}_{\geqslant0}$ is the solution of the following optimization problem:
	\begin{eqnarray}
	f_{L,\gamma,\lambda}(z)=\min_{\omega\in\mathbb{R}_{\geqslant0}}\{\omega\log(\frac{\rvert z\rvert}{\varepsilon}+1)+\frac{\gamma}{2}(\omega-\lambda)^{2}\}.
	\end{eqnarray}
\end{theorem}
\begin{IEEEproof}
	The proof is provided in Appendix B.
\end{IEEEproof}
\begin{theorem}[Vector EMLCP]
	Let $\gamma>0,\varepsilon>0,\omega\in\mathbb{R}_{\geqslant0}^{n},\lambda\in\mathbb{R}_{\geqslant0}^{n}$ and $z\in\mathbb{R}^{n}$. The vector MCP is the solution of the following optimization problem:
	\begin{eqnarray}
	f_{L,\gamma,\lambda}(z)=\min_{\omega\in\mathbb{R}^{n}_{\geqslant0}}\{\|z\|_{L,\omega}+\frac{\gamma}{2}\|\omega-\lambda\|^{2}_{2}\},
	\end{eqnarray}
	where $\|z\|_{L,\omega}$ is defined as 
	\begin{eqnarray}
	\|z\|_{L,\omega}=\sum_{i=1}^{n}\omega_{i}\log(\frac{\rvert z_{i}\rvert}{\varepsilon}+1),\omega_{i}\geqslant0,\nonumber 
	\end{eqnarray}
	and $\{\omega_{i},i=1,2,\dots,n\}$ denote the weights.
\end{theorem}
\begin{IEEEproof}
	The proof is provided in Appendix C.
\end{IEEEproof}
\begin{theorem}[Matrix EMLCP]
	Let $\gamma>0,\varepsilon>0,\Omega\in\mathbb{R}_{\geqslant0}^{m\times n},\Lambda\in\mathbb{R}_{\geqslant0}^{m\times n}$ and $Z\in\mathbb{R}^{m\times n}$. The matrix MLCP is the solution of the following optimization problem:
	\begin{eqnarray}
	f_{L,\gamma,\Lambda}(Z)=\min_{\Omega\in\mathbb{R}^{m\times n}_{\geqslant0}}\{\|Z\|_{L,\Omega}+\frac{\gamma}{2}\|\Omega-\Lambda\|^{2}_{2}\},
	\end{eqnarray}
	where $\|Z\|_{L,\Omega}$ is defined as 
	\begin{eqnarray}
	\|Z\|_{L,\Omega}=\sum_{i=1}^{m}\sum_{j=1}^{n}\Omega_{ij}\log(\frac{\rvert Z_{ij}\rvert}{\varepsilon}+1),\nonumber 
	\end{eqnarray}
	where $\{\Omega_{ij}\geqslant0,i=1,2,\dots,m,j=1,2,\dots,n\}$ denote the weights.
\end{theorem}
\begin{IEEEproof}
	The proof is provided in Appendix D.
\end{IEEEproof}
\begin{theorem}[Tensor EMLCP]
	Let $\gamma>0,\varepsilon>0$, $\mathcal{W},\bar{\lambda}\in\mathbb{R}_{\geqslant0}^{\mathit{I}_{1}\times\mathit{I}_{2}\times\cdots\times\mathit{I}_{N}}$ and $\mathcal{Z}\in\mathbb{R}^{\mathit{I}_{1}\times\mathit{I}_{2}\times\cdots\times\mathit{I}_{N}}$. The tensor MLCP is the solution to the optimization problem:
	\begin{eqnarray}
	f_{L,\gamma,\bar{\lambda}}(\mathcal{Z})=\min_{\mathcal{W}}\{\|\mathcal{Z}\|_{L,\mathcal{W}}+\frac{\gamma}{2}\|\mathcal{W}-\bar{\lambda}\|^{2}_{F}\},\label{tensoremlcp}
	\end{eqnarray}
	where $\|\mathcal{Z}\|_{L,\mathcal{W}}$ is defined as 
	\begin{eqnarray}
	\|\mathcal{Z}\|_{L,\mathcal{W}}=\sum_{i_{1}=1}^{\mathit{I}_{1}}\sum_{i_{2}=1}^{\mathit{I}_{2}}\cdots\sum_{i_{N}=1}^{\mathit{I}_{N}}\mathcal{W}_{t}\log(\frac{\rvert\mathcal{Z}_{t}\rvert}{\varepsilon}+1),
	\end{eqnarray}
	where $\mathcal{W}$ is weight tensor, and $t=i_{1},i_{2},\cdots,i_{N}$.
\end{theorem}
\begin{IEEEproof}
	The proof is provided in Appendix E.
\end{IEEEproof}
\begin{remark}
	As the order of the tensor decreases, tensor EMLCP can degenerate into the form of matrix EMLCP, vector EMLCP, and scalar EMLCP respectively.
\end{remark}
\begin{theorem}[Equivalent weighted matrix $L\gamma$-norm]
	Consider a rank-$r$ matrix $Z\in\mathbb{R}^{m\times n}$ with the SVD: $Z=Udiag(\sigma)V^{T}$, where $\sigma=[\sigma_{1},\sigma_{2},\dots,\sigma_{r}]^{T}$. Let $\Omega,\Lambda\in\mathbb{R}^{r}_{\geqslant0}$, and $\gamma>0,\varepsilon>0$. The matrix $L\gamma$-norm is obtained equivalently as
	\begin{eqnarray}
	\|Z\|_{L,\gamma,\Lambda}=\min_{\Omega}\{\|Z\|_{L,\Omega}+\frac{\gamma}{2}\|\Omega-\Lambda\|^{2}_{2}\},
	\end{eqnarray}
	where $\|Z\|_{L,\Omega}=\sum_{i=1}^{r}\Omega_{i}\log(\frac{\sigma_{i}}{\varepsilon}+1)$.
\end{theorem}
\begin{IEEEproof}
	The proof is provided in Appendix F.
\end{IEEEproof}
\begin{theorem}[Equivalent weighted Tensor $L\gamma$-norm]
	For a third-order tensor $\mathcal{Z}\in\mathbb{R}^{\mathit{I}_{1}\times\mathit{I}_{2}\times\mathit{I}_{3}}$, its SVD is decomposed into $\mathcal{Z}=\mathcal{U}\ast\mathcal{S}\ast\mathcal{V}$, where $\mathcal{S}\in\mathbb{R}^{R\times R\times\mathit{I}_{3}}$ and $R=\min\{\mathit{I}_{1},\mathit{I}_{2}\}$. Let $W,\bar{\Lambda}\in\mathbb{R}^{R\times\mathit{I}_{3}}_{\geqslant0}$, and $\gamma>0,\varepsilon>0$. The weighted tensor $L\gamma$-norm is obtained equivalently as 
	\begin{eqnarray}
	\|\mathcal{Z}\|_{L,\gamma,\bar{\Lambda}}=\min_{W}\{\|\mathcal{Z}\|_{L,W}+\frac{\gamma}{2}\|W-\bar{\Lambda}\|_{F}^{2}\}\label{ewtlgn},
	\end{eqnarray}          
	where
	\begin{eqnarray}
	\|\mathcal{Z}\|_{L,W}:=&&\sum_{i_{3}=1}^{\mathit{I}_{3}}\|\bar{\mathcal{Z}}^{(i_{3})}\|_{L,W_{(:,i_{3})}}\nonumber
	\\=&&\sum_{j=1}^{R}W_{j,i_{3}}\log(\frac{\sigma_{j}(\bar{\mathcal{Z}}^{(i_{3})})}{\varepsilon}+1).\nonumber
	\end{eqnarray} 
\end{theorem}
\begin{IEEEproof}
	The proof is provided in Appendix G.
\end{IEEEproof}
\begin{remark}
	In particular, when the third dimension $\mathit{I}_{3}$ of the third-order tensor $\mathcal{Z}$ is 1, equivalent weighted Tensor $L\gamma$-norm can degenerate into the form of equivalent weighted matrix $L\gamma$-norm.
\end{remark}

\begin{remark}
	Unlike the $l_{1}$ penalty or the nuclear norm penalty, the tensor MLCP (\ref{tensormlcp1}), tensor EMLCP (\ref{tensoremlcp}),  weighted tensor $L\gamma$-norm (\ref{wtgnmlcp}), and equivalent weighted tensor $L\gamma$-norm (\ref{ewtlgn}) do not satisfy the triangle inequality. Some vital properties of the tensor EMLCP and equivalent weighted tensor $L\gamma$-norm are given below.
\end{remark}

\begin{proposition}
	The tensor EMLCP $f_{L,\gamma,\bar{\lambda}}(\mathcal{Z})$ is defined in (\ref{tensoremlcp}) satisfies the following properties:
	\\\textbf{(a) Non-negativity}: The tensor EMLCP is non-negative, i.e., $f_{L,\gamma,\bar{\lambda}}(\mathcal{Z})\geqslant0$. The equality holds if and only if $\mathcal{Z}$ is the null tensor. 
	\\\textbf{(b) Concavity}:
	$f_{L,\gamma,\bar{\lambda}}(\mathcal{Z})$ is concave in the modulus of the elements of $\mathcal{Z}$.
	\\\textbf{(c) Boundedness}: The tensor EMLCP is upper-bounded by the weighted Logarithmic norm, i.e., $f_{L,\gamma,\bar{\lambda}}(\mathcal{Z})\leqslant\|\mathcal{Z}\|_{L,\mathcal{W}}.$
	\\\textbf{(d) Asymptotic $l_{1}$ property}: The tensor EMLCP approaches the weighted Logarithmic norm asymptotically, i.e., $\lim\limits_{\gamma\to\infty}f_{L,\gamma,\bar{\lambda}}(\mathcal{Z})=\|\mathcal{Z}\|_{L,\mathcal{W}}$
\end{proposition}
\begin{IEEEproof}
	The proof is provided in Appendix H.
\end{IEEEproof}
\begin{proposition}
	The equivalent weighted tensor $L\gamma$-norm is defined in (\ref{ewtlgn}) satisfies the following properties:
	\\\textbf{(a) Non-negativity}: The equivalent weighted tensor $L\gamma$-norm is non-negative, i.e., $\|\mathcal{Z}\|_{L,\gamma,\bar{\Lambda}}\geqslant0$. The equality holds if and only if $\mathcal{Z}$ is the null tensor. 
	\\\textbf{(b) Concavity}:
	$\|\mathcal{Z}\|_{L,\gamma,\bar{\Lambda}}$ is concave in the modulus of the elements of $\mathcal{Z}$.
	\\\textbf{(c) Boundedness}: The equivalent weighted tensor $L\gamma$-norm is upper-bounded by the weighted Logarithmic norm, i.e., $\|\mathcal{Z}\|_{L,\gamma,\bar{\Lambda}}\leqslant\|\mathcal{Z}\|_{L,W}.$
	\\\textbf{(d) Asymptotic nuclear norm property}: The equivalent weighted tensor $L\gamma$-norm approaches the weighted Logarithmic norm asymptotically, i.e., $\lim\limits_{\gamma\to\infty}\|\mathcal{Z}\|_{L,\gamma,\bar{\Lambda}}=\|\mathcal{Z}\|_{L,W}.$
	\\\textbf{(e) Unitary invariance}: The equivalent weighted tensor $L\gamma$-norm is unitary invariant, i.e., $\|\mathcal{U}\ast\mathcal{Z}\ast\mathcal{V}\|_{L,\gamma,\bar{\Lambda}}=\|\mathcal{Z}\|_{L,\gamma,\bar{\Lambda}}$, for unitary tensor $\mathcal{U}$ and $\mathcal{V}$.
\end{proposition}
\begin{IEEEproof}
	The proof is provided in Appendix I.
\end{IEEEproof}
\begin{theorem}[Proximal operator for equivalent weighted tensor $L\gamma$-norm]
	Consider equivalent weighted tensor $L\gamma$-norm given in (\ref{ewtlgn}). Its proximal operator denoted by $\mathit{S}_{L,\gamma,\bar{\Lambda}}:\mathbb{R}^{\mathit{I}_{1}\times\mathit{I}_{2}\times\mathit{I}_{3}}\to\mathbb{R}^{\mathit{I}_{1}\times\mathit{I}_{2}\times\mathit{I}_{3}}$, $W,\bar{\Lambda}\in\mathbb{R}^{R\times\mathit{I}_{3}}_{\geqslant0}$, and $\gamma>0,\varepsilon>0$, $R=\min\{\mathit{I}_{1},\mathit{I}_{2}\}$ and defined as follows:
	\begin{eqnarray}
	\mathit{S}_{L,\gamma,\bar{\Lambda}}(\mathcal{Y})=\arg\min_{\mathcal{L}}\{\frac{\rho}{2}\|\mathcal{L}-\mathcal{Y}\|_{F}^{2}+\|\mathcal{L}\|_{L,\gamma,\bar{\Lambda}}\},\label{prox}
	\end{eqnarray}
	is given by
	\begin{eqnarray}
	\mathit{S}_{L,\gamma,\bar{\Lambda}}=\left\{\begin{array}{l}
	W_{j,i}=\max\{\bar{\Lambda}_{j,i}-\frac{\log(\frac{\sigma_{j}(\bar{\mathcal{L}}^{(i)})}{\varepsilon}+1)}{\gamma},0\},
	\\\mathcal{L}=\mathcal{U}\ast\mathcal{S}_{1}\ast\mathcal{V}^{H},
	\end{array}
	\right.\label{opewtgn}
	\end{eqnarray}
	
	where $\mathcal{U}$ and $\mathcal{V}$ are derived from the t-SVD of $\mathcal{Y}=\mathcal{U}\ast\mathcal{S}_{2}\ast\mathcal{V}^{H}$. More importantly, the $i$th front slice of DFT of $\mathcal{S}_{1}$ and $\mathcal{S}_{2}$, i.e., $\bar{\mathcal{S}}^{(i)}_{1}=\sigma(\bar{\mathcal{L}}^{(i)})$ and $\bar{\mathcal{S}}^{(i)}_{2}=\sigma(\bar{\mathcal{Y}}^{(i)})$, has the following relationship
	\begin{eqnarray}
	\sigma_{j}(\bar{\mathcal{L}}^{(i)})=\left\{\begin{array}{l}
	0,\qquad if\quad\sigma_{j}(\bar{\mathcal{Y}}^{(i)})\leqslant 2\sqrt{\alpha}-\varepsilon,
	\\\frac{l_{1}+l_{2}}{2},\quad if\quad\sigma_{j}(\bar{\mathcal{Y}}^{(i)})> 2\sqrt{\alpha}-\varepsilon,
	\end{array}
	\right.
	\end{eqnarray}
	where $l_{1}=\sigma_{j}(\bar{\mathcal{Y}}^{(i)})-\varepsilon,l_{2}=\sqrt{(\sigma_{j}(\bar{\mathcal{Y}}^{(i)})+\varepsilon)^{2}-4\alpha}$, $\alpha=\frac{W_{j,i}}{\rho}$.
	\label{thEWTLGN}
\end{theorem}
\begin{IEEEproof}
	The proof is provided in Appendix J.
\end{IEEEproof}
\section{EMLCP-based models and solving algorithms}

In this section, we apply the EMLCP to low rank tensor completion (LRTC) and tensor robust principal component analysis (TRPCA) and propose the EMLCP-based models with proximal alternating linearized minimization algorithms.
\subsection{EMLCP-based LRTC model}
\begin{algorithm}[t]
	\caption{EMLCPTC} 
	\hspace*{0.02in} {\bf Input:} 
	An incomplete tensor $\mathcal{T}$, the index set of the known elements $\Omega$, convergence criteria $\epsilon$, maximum iteration number $K$. \\
	\hspace*{0.02in} {\bf Initialization:} 
	$\mathcal{Z}^{0}=\mathcal{T}_{\Omega}$, $\mathcal{M}_{k_{1}k_{2}}^{0}=\mathcal{X}^{0}$, $\mu_{k_{1}k_{2}}^{0}>0$, $\rho>0$,$\tau>1$.
	\begin{algorithmic}[1]
		\While{not converged and $k<K$} 
		\State Updating $W_{k_{1}k_{2}}^{k}$ via (\ref{UPW1});
		\State Updating $\mathcal{M}_{k_{1}k_{2}}^{k}$ via (\ref{UPM1});
		\State Updating $\mathcal{Z}^{k}$ via (\ref{UPX1});
		\State Updating the multipliers $\mathcal{Q}_{k_{1}k_{2}}^{k}$ via (\ref{UPQ1});
		\State $\mu_{k_{1}k_{2}}^{k}=\tau\mu_{k_{1}k_{2}}^{k-1}$, $\rho=\tau\rho$ $k=k+1$;
		\State Check the convergence conditions $\|\mathcal{Z}^{k+1}-\mathcal{Z}^{k}\|_{\infty}\leq\epsilon$.
		\EndWhile
		\State \Return $\mathcal{Z}^{k+1}$.
	\end{algorithmic}
	\hspace*{0.02in} {\bf Output:} 
	Completed tensor $\mathcal{Z}=\mathcal{Z}^{k+1}$.
	\label{TC}\end{algorithm}
Tensor completion aims at estimating the missing elements from an incomplete observation tensor. Considering an $N$-order tensor $\mathcal{Z}\in\mathbb{R}^{\mathit{I}_{1}\times\mathit{I}_{2}\times\cdots\times\mathit{I}_{N}}$, the proposed EMLCP-based LRTC model is formulated as follow
\begin{eqnarray}
\min_{\mathcal{Z},W}\sum_{1\leqslant k_{1}<k_{2}\leqslant N}&\beta_{k_{1}k_{2}}(\|\mathcal{Z}_{(k_{1}k_{2})}\|_{L,W}+\frac{\gamma}{2}\|W_{(k_{1}k_{2})}-\bar{\Lambda}\|_{F}^{2})\nonumber
\\&+\Phi_{\mathbb{G}}(\mathcal{Z})\label{FTC1}
\end{eqnarray}
where $\mathcal{Z}$ is the reconstructed tensor and $\mathcal{T}$ is the observed tensor, $\Omega$ is the index set for the known entries, and $\mathcal{P}_{\Omega}(\mathcal{Z})$ is a projection
operator that keeps the entries of $\mathcal{Z}$ in $\Omega$ and sets all others to zero, $\beta_{k_{1}k_{2}}\geqslant0$ $(1\leqslant k_{1}<k_{2}\leqslant N,k_{1},k_{2}\in\mathbb{Z})$ and $\sum_{1\leqslant k_{1}< k_{2}\leqslant N}\beta_{k_{1}k_{2}}=1$. Let
\begin{eqnarray}
\Phi_{\mathbb{G}}(\mathcal{Z}):=\left\{\begin{array}{l}
0,\qquad if\quad\mathcal{Z}\in\mathbb{G},
\\\infty,\qquad otherwise
\end{array}
\right.
\end{eqnarray}
where $\mathbb{G}:=\{\mathcal{Z}\in\mathbb{R}^{\mathit{I}_{1}\times\mathit{I}_{2}\times\cdots\times\mathit{I}_{N}},\mathcal{P}_{\Omega}(\mathcal{Z}-\mathcal{T})=0\}$.

Next, we exploit the PALM to solve (\ref{FTC1}). We first introduce auxiliary variables $\mathcal{M}_{k_{1}k_{2}}$, and then rewrite (\ref{FTC1}) as the following equivalent constrained problem:
\begin{eqnarray}
\min_{\mathcal{Z},W}&&\sum_{1\leqslant k_{1}< k_{2}\leqslant\nonumber N}\beta_{k_{1}k_{2}}(\|\mathcal{M}_{(k_{1}k_{2})}\|_{L,W}+\frac{\gamma}{2}\|W_{(k_{1}k_{2})}-\bar{\Lambda}\|_{F}^{2})
\\&&+\Phi_{\mathbb{G}}(\mathcal{Z})\label{FTC3}
\\&&s.t.\quad \mathcal{Z}=\mathcal{M}_{k_{1}k_{2}},1\leqslant k_{1}< k_{2}\leqslant N,k_{1},k_{2}\in\mathbb{Z}.\nonumber
\end{eqnarray}
The augmented Lagrangian function of (\ref{FTC3}) can be expressed in the following concise form:
\begin{eqnarray}
&&LAG(\mathcal{Z},\mathcal{M},W,\bar{\Lambda},\mathcal{Q})=\nonumber
\\&&\sum_{1\leqslant k_{1}< k_{2}\leqslant N}\beta_{k_{1}k_{2}}(\|\mathcal{M}_{(k_{1}k_{2})}\|_{L,W}+\frac{\gamma}{2}\|W_{(k_{1}k_{2})}-\bar{\Lambda}\|_{F}^{2})\nonumber
\\&&+\Phi_{\mathbb{G}}(\mathcal{Z})+\frac{\mu_{k_{1}k_{2}}}{2}\|\mathcal{Z}-\mathcal{M}_{k_{1}k_{2}}+\frac{\mathcal{Q}_{k_{1}k_{2}}}{\mu_{k_{1}k_{2}}}\|_{F}^{2},\nonumber
\\&&=H_{1}(\mathcal{M},W)+H_{2}(W,\bar{\Lambda})+\Phi_{\mathbb{G}}(\mathcal{Z})+H_{3}(\mathcal{M},\mathcal{Z}),\label{TCAG}
\end{eqnarray}
where $\mathcal{Q}_{k_{1}k_{2}}(1\leqslant k_{1}\leqslant k_{2}\leqslant N)$ are the Lagrange multipliers, $\mu_{k_{1}k_{2}}$ are positive scalars. For the sake of convenience, we denote the variable updated by the iteration as $(\cdot)^{+}$, the last iteration result as $(\cdot)^{*}$, and omit the specific number of iterations. With the proximal linearization of each subproblem, the PALM algorithm on the four blocks $(\mathcal{Z},\mathcal{M},W,\bar{\Lambda})$ for solving (\ref{FTC3}) yields the iteration scheme alternatingly as follows:
\begin{eqnarray}
\left\{\begin{array}{l}
W^{+}=\min_{W} H_{1}(W)+H_{2}(W)+\frac{\rho^{*}}{2}\|W-W^{*}\|_{F}^{2},
\\\mathcal{M}^{+}=\min_{\mathcal{M}} H_{1}(\mathcal{M})+\langle \mathcal{M}-\mathcal{M}^{*},\nabla H_{3}(\mathcal{M}^{*}) \rangle
\\ +\frac{\rho_{1}^{*}}{2}\|\mathcal{M}-\mathcal{M}^{*}\|_{F}^{2},
\\\bar{\Lambda}^{+}=\min_{\bar{\Lambda}} H_{2}(\bar{\Lambda})+\frac{\rho^{*}}{2}\|\bar{\Lambda}-\bar{\Lambda}^{*}\|_{F}^{2},
\\\mathcal{Z}^{+}=\min_{\mathcal{Z}} \Phi_{\mathbb{G}}(\mathcal{Z})+H_{3}(\mathcal{Z})+\frac{\rho^{*}}{2}\|\mathcal{Z}-\mathcal{Z}^{*}\|_{F}^{2},
\end{array}
\right.\label{PA}
\end{eqnarray}

From Theorem $\ref{thEWTLGN}$, the updates of $W$ and $\mathcal{M}$ are as follows:
\begin{eqnarray}
W_{j,i}^{+}=\max(\frac{\gamma\bar{\Lambda}^{*}_{j,i}+\rho^{*} W_{j,i}^{*}-log(\frac{\sigma_{j}(\mathcal{M}^{*(i)})}{\varepsilon}+1)}{\gamma+\rho^{*}},0),\label{UPW1}
\end{eqnarray}
\begin{eqnarray}
\mathcal{M}^{+}=\mathcal{U}\ast\mathcal{S}_{1}\ast\mathcal{V}^{H},\label{UPM1}
\end{eqnarray}
where $\mathcal{U}$ and $\mathcal{V}$ are derived from the t-SVD of $\mathcal{M}^{*}+\frac{\mu\mathcal{Z}^{*}+\mathcal{Q}^{*}-\mu\mathcal{M}^{*}}{\rho_{1}^{*}}=\mathcal{U}\ast\mathcal{S}_{2}\ast\mathcal{V}^{H}$. The relationship between $\mathcal{S}_{1}$ and $\mathcal{S}_{2}$ is given by Theorem $\ref{thEWTLGN}$.

The update for $\bar{\Lambda}$ turns out to be straightforward:
\begin{eqnarray}
\bar{\Lambda}^{+}&&=\min_{\bar{\Lambda}}\frac{\gamma}{2}\|W^{+}-\bar{\Lambda}\|_{F}^{2}+\frac{\rho^{*}}{2}\|\bar{\Lambda}-\bar{\Lambda}^{*}\|_{F}^{2}\nonumber
\\&&=\frac{\gamma W^{+}+\rho^{*}\bar{\Lambda}^{*}}{\gamma+\rho^{*}}.\label{UPLA1}
\end{eqnarray}

Fixed $W_{k_{1}k_{2}}$, $\mathcal{M}_{k_{1}k_{2}}$, $\bar{\Lambda}_{k_{1}k_{2}}$ and $\mathcal{Q}_{k_{1}k_{2}}$, the minimization problem of $\mathcal{Z}$ is as follows:
\begin{eqnarray}
\min_{\mathcal{Z}}\sum_{1\leqslant k_{1}< k_{2}\leqslant N}\frac{\mu_{k_{1}k_{2}}}{2}\|\mathcal{Z}-\mathcal{M}^{+}_{k_{1}k_{2}}+\frac{\mathcal{Q}^{*}_{k_{1}k_{2}}}{\mu_{k_{1}k_{2}}}\|_{F}^{2}\nonumber
\\+\Phi_{\mathbb{G}}(\mathcal{Z})+\frac{\rho^{*}}{2}\|\mathcal{Z}-\mathcal{Z}^{*}\|_{F}^{2}.\label{FFXX1}
\end{eqnarray}
The closed form of $\mathcal{Z}$ can be derived by setting the derivative of (\ref{FFXX1}) to zero. We can now update $\mathcal{Z}$ by the following equation:
\begin{eqnarray}
&&\mathcal{Z}^{+}=\mathcal{P}_{\Omega}(\mathcal{T})\nonumber
\\&&+\mathcal{P}_{\Omega^{c}}(\dfrac{\sum_{1\leqslant k_{1}<k_{2}\leqslant N}\mu_{k_{1}k_{2}}\mathcal{M}^{*}_{k_{1}k_{2}}-\mathcal{Q}^{*}_{k_{1}k_{2}}+\rho^{*}\mathcal{Z}^{*}}{\sum_{1\leqslant k_{1}< k_{2}\leqslant N}\mu_{k_{1}k_{2}}+\rho^{*}}).\label{UPX1}
\end{eqnarray}
Finally, multipliers $\mathcal{Q}_{k_{1}k_{2}}$ are updated as follows:
\begin{eqnarray}
\mathcal{Q}_{k_{1}k_{2}}^{+}=\mathcal{Q}^{*}_{k_{1}k_{2}}+\mu_{k_{1}k_{2}}(\mathcal{Z}^{+}-\mathcal{M}^{+}_{k_{1}k_{2}}).\label{UPQ1}
\end{eqnarray}

The EMLCP-based LRTC model computation is given in Algorithm \ref{TC}. The main per-iteration cost lies in the update of $\mathcal{M}_{k_{1}k_{2}}$, which requires computing t-SVD. The per-iteration complexity is $O(LE(\sum_{1\leqslant k_{1}< k_{2}\leqslant N}[log(le_{k_{1}k_{2}})+\min(\mathit{I}_{k_{1}},\mathit{I}_{k_{2}})]))$, where $LE=\prod_{i=1}^{N}\mathit{I}_{i}$ and $le_{k_{1}k_{2}}=LE/(\mathit{I}_{k_{1}}\mathit{I}_{k_{2}})$.

\subsection{EMLCP-based TRPCA model}
Tensor robust PCA (TRPCA) aims to recover the tensor from grossly corrupted observations. Using the proposed EMLCP, we can get the following EMLCP-based TRPCA model:
\begin{eqnarray}
\min_{\mathcal{L},\mathcal{E},\mathcal{N}}\sum_{1\leqslant k_{1}<k_{2}\leqslant N}\beta_{k_{1}k_{2}}(\|\mathcal{L}_{(k_{1}k_{2})}\|_{L,W}+\frac{\gamma}{2}\|W_{(k_{1}k_{2})}-\bar{\Lambda}\|_{F}^{2})\nonumber
\\+\tau_{1}\|\mathcal{E}\|_{1}+\tau_{2}\|\mathcal{N}\|_{F}\quad s.t.\quad \mathcal{T}=\mathcal{L}+\mathcal{E}+\mathcal{N},\label{FPCA1}
\end{eqnarray}
where $\mathcal{T}$ is the corrupted observation tensor, $\mathcal{L}$ is the low-rank component, $\mathcal{E}$ is the sparse noise component, $\mathcal{N}$ is the Gaussian noise component, and $\tau_{1},\tau_{2}$ are tuning parameters compromising $\mathcal{L}$, $\mathcal{E}$ and $\mathcal{N}$. Similarly, we introduce auxiliary variables $\mathcal{G}_{k_{1}k_{2}}$, and then rewrite (\ref{FPCA1}) as the following equivalent constrained problem:
\begin{eqnarray}
\min_{\mathcal{L},\mathcal{E},\mathcal{N}}&&\sum_{1\leqslant k_{1}< k_{2}\leqslant N}\beta_{k_{1}k_{2}}(\|\mathcal{G}_{(k_{1}k_{2})}\|_{L,W}+\frac{\gamma}{2}\|W_{(k_{1}k_{2})}-\bar{\Lambda}\|_{F}^{2})\nonumber\\&&+\tau_{1}\|\mathcal{E}\|_{1}+\tau_{2}\|\mathcal{N}\|_{F}\nonumber
\\&&s.t.\mathcal{T}=\mathcal{L}+\mathcal{E}+\mathcal{N},\nonumber
\\&&\quad \mathcal{L}=\mathcal{G}_{k_{1}k_{2}},1\leqslant k_{1}< k_{2}\leqslant N,k_{1},k_{2}\in\mathbb{Z}.
\label{FPCA2}\end{eqnarray}
\begin{algorithm}[t]
	\caption{EMLCPTPRCA} 
	\hspace*{0.02in} {\bf Input:} 
	The corrupted observation tensor $\mathcal{T}$, convergence criteria $\epsilon$, maximum iteration number $K$. \\
	\hspace*{0.02in} {\bf Initialization:} 
	$\mathcal{L}^{0}=\mathcal{T}$, $\mathcal{G}_{k_{1}k_{2}}^{0}=\mathcal{L}^{0}$, $\mu_{k_{1}k_{2}}^{0}>0$, $\rho^{0}>0$, $\tau^{0}>0$, $\upsilon>1$.
	\begin{algorithmic}[1]
		\While{not converged and $k<K$} 
		\State Updating $W_{k_{1}k_{2}}^{k}$ via (\ref{UPW2});
		\State Updating $\mathcal{G}_{k_{1}k_{2}}^{k}$ via (\ref{UPG2});
		\State Updating $\bar{\Lambda}_{k_{1}k_{2}}^{k}$ via (\ref{UPLA2});
		\State Updating $\mathcal{L}^{k}$ via (\ref{UPL2});
		\State Updating $\mathcal{E}^{k}$ via (\ref{UPE2});
		\State Updating $\mathcal{N}^{k}$ via (\ref{UPN2});
		\State Updating the multipliers $\mathcal{R}_{k_{1}k_{2}}^{k}$ and $\mathcal{F}^{k}$ via (\ref{UPM21});
		\State $\mu_{k_{1}k_{2}}^{k}=\upsilon\mu_{k_{1}k_{2}}^{k-1}$, $\rho^{k}=\upsilon\rho^{k-1}$, $\tau^{k}=\upsilon\tau^{k-1}$ $k=k+1$;
		\State Check the convergence conditions $\|\mathcal{L}^{k+1}-\mathcal{L}^{k}\|_{\infty}\leq\epsilon$.
		\EndWhile
		\State \Return $\mathcal{L}^{k+1}$, $\mathcal{E}^{k+1}$ and $\mathcal{N}^{k+1}$.
	\end{algorithmic}
	\hspace*{0.02in} {\bf Output:} 
	$\mathcal{L}$ and $\mathcal{E}$.
	\label{TPRCA}\end{algorithm}
The augmented Lagrangian function of (\ref{FPCA2}) can be expressed in the following concrete form:
\begin{eqnarray}
&&LAG(\mathcal{L},\mathcal{G},W,\bar{\Lambda},\mathcal{R},\mathcal{E},\mathcal{N},\mathcal{F})=\nonumber
\\&&\sum_{1\leqslant k_{1}< k_{2}\leqslant N}\beta_{k_{1}k_{2}}(\|\mathcal{G}_{(k_{1}k_{2})}\|_{L,W}+\frac{\gamma}{2}\|W_{(k_{1}k_{2})}-\bar{\Lambda}\|_{F}^{2})\nonumber
\\&&+\tau_{1}\|\mathcal{E}\|_{1}+\tau_{2}\|\mathcal{N}\|_{F}+\frac{\mu_{k_{1}k_{2}}}{2}\|\mathcal{L}-\mathcal{G}_{k_{1}k_{2}}+\frac{\mathcal{R}_{k_{1}k_{2}}}{\mu_{k_{1}k_{2}}}\|_{F}^{2}\nonumber
\\&&+\frac{\tau}{2}\|\mathcal{T}-\mathcal{L}-\mathcal{E}-\mathcal{N}+\frac{\mathcal{F}}{\tau}\|_{F}^{2}\nonumber
\\&&=H_{1}(\mathcal{G},W)+H_{2}(W,\bar{\Lambda})+H_{3}(\mathcal{E})\nonumber
\\&&+H_{4}(\mathcal{N})+H_{5}(\mathcal{L},\mathcal{G})+H_{6}(\mathcal{L},\mathcal{E},\mathcal{N})
\end{eqnarray}
where $\mathcal{F}$ and ${R}_{k_{1}k_{2}}(1\leqslant k_{1}< k_{2}\leqslant N)$ are the Lagrange multipliers, $\mu_{k_{1}k_{2}}$, $\tau_{1}$, $\tau_{2}$ and $\tau$ are positive scalars. Similar to EMLCP-based LRTC model, we denote the variable updated by the iteration as $(\cdot)^{+}$, the last iteration result as $(\cdot)^{*}$, and omit the specific number of iterations. With the proximal linearization of each subproblem, the PALM algorithm on the six blocks $(\mathcal{L},\mathcal{G},W,\bar{\Lambda},\mathcal{E},\mathcal{N})$ for solving (\ref{FTC3}) yields the iteration scheme alternatingly as follows:
\begin{eqnarray}
\left\{\begin{array}{l}
W^{+}=\min_{W} H_{1}(W)+H_{2}(W)+\frac{\rho^{*}}{2}\|W-W^{*}\|_{F}^{2},
\\\mathcal{G}^{+}=\min_{\mathcal{G}} H_{1}(\mathcal{G})+\langle \mathcal{G}-\mathcal{G}^{*},\nabla H_{5}(\mathcal{G}^{*}) \rangle
\\ +\frac{\rho_{1}^{*}}{2}\|\mathcal{G}-\mathcal{G}^{*}\|_{F}^{2},
\\\bar{\Lambda}^{+}=\min_{\bar{\Lambda}} H_{2}(\bar{\Lambda})+\frac{\rho^{*}}{2}\|\bar{\Lambda}-\bar{\Lambda}^{*}\|_{F}^{2},
\\\mathcal{L}^{+}=\min_{\mathcal{L}} H_{5}(\mathcal{L})+H_{6}(\mathcal{L})+\frac{\rho^{*}}{2}\|\mathcal{L}-\mathcal{L}^{*}\|_{F}^{2},
\\\mathcal{E}^{+}=\min_{\mathcal{E}} H_{3}(\mathcal{E})+H_{6}(\mathcal{E})+\frac{\rho^{*}}{2}\|\mathcal{E}-\mathcal{E}^{*}\|_{F}^{2},
\\\mathcal{N}^{+}=\min_{\mathcal{N}} H_{4}(\mathcal{N})+H_{6}(\mathcal{N})+\frac{\rho^{*}}{2}\|\mathcal{N}-\mathcal{N}^{*}\|_{F}^{2},
\end{array}
\right.\label{PA2}
\end{eqnarray}
Based on Theorem $\ref{thEWTLGN}$, $W$ and $\mathcal{G}$ are updated as follows:
\begin{eqnarray}
W_{j,i}^{+}=\max(\frac{\gamma\bar{\Lambda}^{*}_{j,i}+\rho W_{j,i}^{*}-log(\frac{\sigma_{j}(\mathcal{G}^{*(i)})}{\varepsilon}+1)}{\gamma+\rho^{*}},0),\label{UPW2}
\end{eqnarray}
\begin{eqnarray}
\mathcal{G}^{+}=\mathcal{U}\ast\mathcal{S}_{1}\ast\mathcal{V}^{H},\label{UPG2}
\end{eqnarray}
where $\mathcal{U}$ and $\mathcal{V}$ are derived from the t-SVD of $\mathcal{G}^{*}+\frac{\mu\mathcal{L}^{*}+\mathcal{R}^{*}-\mu\mathcal{G}^{*}}{\rho_{1}^{*}}=\mathcal{U}\ast\mathcal{S}_{2}\ast\mathcal{V}^{H}$. The relationship between $\mathcal{S}_{1}$ and $\mathcal{S}_{2}$ is given by Theorem $\ref{thEWTLGN}$.

The update for $\bar{\Lambda}$ turns out to be straightforward:
\begin{eqnarray}
\bar{\Lambda}^{+}=\frac{\gamma W^{+}+\rho^{*}\bar{\Lambda}^{*}}{\gamma+\rho^{*}}.\label{UPLA2}
\end{eqnarray}

Fixed $\mathcal{G}_{k_{1}k_{2}}$, $\mathcal{E}$, $\mathcal{N}$, $\mathcal{R}_{k_{1}k_{2}}$ and $\mathcal{F}$, the minimization problem $\mathcal{L}$ is converted into the following form:
\begin{eqnarray}
&&\min_{\mathcal{L}}\sum_{1\leqslant l_{1}< l_{2}\leqslant N}\beta_{k_{1}k_{2}}\frac{\mu_{k_{1}k_{2}}}{2}\|\mathcal{L}-\mathcal{G}^{+}_{k_{1}k_{2}}+\frac{\mathcal{R}^{*}_{k_{1}k_{2}}}{\mu_{k_{1}k_{2}}}\|_{F}^{2}\nonumber
\\&&+\frac{\tau}{2}\|\mathcal{T}-\mathcal{L}-\mathcal{E}^{*}-\mathcal{N}^{*}+\frac{\mathcal{F}^{*}}{\tau}\|_{F}^{2}+\frac{\rho^{*}}{2}\|\mathcal{L}-\mathcal{L}^{*}\|_{F}^{2}.\label{FFLL1}
\end{eqnarray}
The closed form of $\mathcal{L}$ can be derived by setting the derivative of (\ref{FFLL1}) to zero. We can now update $\mathcal{L}$ by the following equation:
\begin{eqnarray}
\mathcal{L}^{+}=\frac{\mathcal{S}}{\sum\mu_{k_{1}k_{2}}+\tau+\rho^{*}},\label{UPL2}
\end{eqnarray}
where $\mathcal{S}=\sum\mu_{k_{1}k_{2}}\mathcal{G}^{+}_{k_{1}k_{2}}-\mathcal{R}^{*}_{k_{1}k_{2}}+\tau(\mathcal{T}-\mathcal{E}^{*}-\mathcal{N}^{*})+\mathcal{F}^{*}+\rho^{*}\mathcal{L}^{*}$.

Now, let's solve $\mathcal{E}$. The minimization problem of $\mathcal{E}$ is as follows:
\begin{eqnarray}
\min_{\mathcal{E}}\tau_{1}\|\mathcal{E}\|_{1}+\frac{\rho^{*}}{2}\|\mathcal{E}-\mathcal{E}^{*}\|_{F}^{2}\nonumber
\\+\frac{\tau}{2}\|\mathcal{T}-\mathcal{L}^{+}-\mathcal{E}-\mathcal{N}^{*}+\frac{\mathcal{F}^{*}}{\tau}\|_{F}^{2}.\label{FFEE1}
\end{eqnarray}
Problem (\ref{FFEE1}) has the following closed-form solution:
\begin{eqnarray}
\mathcal{E}^{+}=S_{\frac{\tau_{1}}{\tau+\rho^{*}}}(\dfrac{\tau(\mathcal{T}-\mathcal{L}^{+}+\frac{\mathcal{F}^{*}}{\tau})+\rho^{*}\mathcal{E}^{*}}{\tau+\rho^{*}}),\label{UPE2}
\end{eqnarray}
where $S_{\lambda}(\cdot)$ is the soft thresholding operator \cite{2011733}: 
\begin{eqnarray}
S_{\lambda}(x)=\left\{\begin{array}{c}
0,\quad if\quad|x|\leqslant \lambda,
\\sign(x)(|x|-\lambda),\quad if\quad|x|> \lambda
\end{array}
\right.
\end{eqnarray}
The minimization problem of $\mathcal{N}$ is as follows:
\begin{eqnarray}
\min_{\mathcal{N}}\tau_{2}\|\mathcal{N}\|_{F}^{2}+\frac{\tau}{2}\|\mathcal{T}-\mathcal{L}^{+}-\mathcal{E}^{+}-\mathcal{N}+\frac{\mathcal{F}^{*}}{\tau}\|_{F}^{2}+\frac{\rho^{*}}{2}\|\mathcal{N}-\mathcal{N}^{*}\|_{F}^{2}.\nonumber
\end{eqnarray}
We update $\mathcal{N}$ by the following equation:
\begin{eqnarray}
\mathcal{N}^{+}=\frac{\tau(\mathcal{T}-\mathcal{L}^{+}-\mathcal{E}^{+})+\mathcal{F}^{*}+\rho\mathcal{N}^{*}}{2\tau_{2}+\tau+\rho^{*}}.\label{UPN2}
\end{eqnarray}
Finally, multipliers $\mathcal{R}_{k_{1}k_{2}}$ and $\mathcal{F}$ are updated according to the following formula:
\begin{eqnarray}
\left\{\begin{array}{l}
\mathcal{R}_{k_{1}k_{2}}^{+}=\mathcal{R}^{*}_{k_{1}k_{2}}+\mu_{k_{1}k_{2}}(\mathcal{L}^{+}-\mathcal{G}^{+}_{k_{1}k_{2}});
\\\mathcal{F}^{+}=\mathcal{F}^{*}+\tau(\mathcal{T}-\mathcal{L}^{+}-\mathcal{E}^{+}-\mathcal{N}^{+}).
\end{array}
\right.
\label{UPM21}\end{eqnarray}
EMLCP-based TPRCA model computation is given in Algorithm \ref{TPRCA}. The main per-iteration cost lies in the update of $\mathcal{G}_{k_{1}k_{2}}$, which requires computing SVD and t-SVD . The per-iteration complexity is $O(LE(\sum_{1\leqslant k_{1}< k_{2}\leqslant N}[log(le_{k_{1}k_{2}})+\min(\mathit{I}_{k_{1}},\mathit{I}_{k_{2}})]))$, where $LE=\prod_{i=1}^{N}\mathit{I}_{i}$ and $le_{k_{1}k_{2}}=LE/(\mathit{I}_{k_{1}}\mathit{I}_{k_{2}})$.
\section{Convergence analysis}
In this section, the convergence of PALM is established under some mild conditions, which is mainly based on the framework in \cite{bolte2014proximal}.

\begin{theorem} Suppose that $\rho_{1}=\gamma_{1}\mu$ with $\gamma_{1}>1$. Let the sequence $\{(\mathcal{Z},\mathcal{M},W,\bar{\Lambda})\}$ be generated by Algorithm \ref{TC}. Then,
	\\ (a) any accumulation point of the sequence $\{(\mathcal{Z}^{k},\mathcal{M}^{k},W^{k},\bar{\Lambda}^{k})\}$ is a critical point of (\ref{FTC1}).
	\\ (b) if $H_{1}$ is KL functions, the sequence $\{(\mathcal{Z}^{k},\mathcal{M}^{k},W^{k},\bar{\Lambda}^{k})\}$ converges to a critical point of (\ref{FTC1}).
\end{theorem}
\begin{IEEEproof}
	First, in the solution process, only the non-convex function $H_{1}(\mathcal{M})$ is included in the iteration of $\mathcal{M}$, and the updates of other elements are solved by convex functions. The variables solved by the convex function are strictly descending, hence we get the following inequality: 
	\begin{eqnarray}
    H_{1}(W^{k+1})+H_{2}(W^{k+1})+\frac{\rho^{k}}{2}\|W^{k+1}-W^{k}\|_{F}^{2}\nonumber
    \\\leqslant H_{1}(W^{k})+H_{2}(W^{k})+\frac{\rho^{k}}{2}\|W^{k}-W^{k}\|_{F}^{2},
    	\end{eqnarray}
    	\begin{eqnarray}
    H_{2}(\bar{\Lambda}^{k+1})+\frac{\rho^{k}}{2}\|\bar{\Lambda}^{k+1}-\bar{\Lambda}^{k}\|_{F}^{2}\leqslant H_{2}(\bar{\Lambda}^{k}),
    	\end{eqnarray}
	\begin{eqnarray}
	H_{3}(\mathcal{Z}^{k+1})+\frac{\rho^{k}}{2}\|\mathcal{Z}^{k+1}-\mathcal{Z}^{k}\|_{F}^{2}\leqslant H_{3}(\mathcal{Z}^{k}),
	\end{eqnarray}
	By the definition of $H_{3}(\mathcal{M},\mathcal{Z})$ in (\ref{TCAG}), the gradients of $H_{3}$ with respect to $\mathcal{M}$ and $\mathcal{Z}$, respectively, are 
	\begin{eqnarray}
	\nabla_{\mathcal{M}}H_{3}(\mathcal{M},\mathcal{Z})=\mu(\mathcal{M}-\mathcal{Z}-\frac{\mathcal{Q}}{\mu}),
	\\\nabla_{\mathcal{Z}}H_{3}(\mathcal{M},\mathcal{Z})=\mu(\mathcal{Z}-\mathcal{M}+\frac{\mathcal{Q}}{\mu}).
	\end{eqnarray}
	For any fixed $\mathcal{Z}$, we obtain that
	\begin{eqnarray}
	&&\|\nabla_{\mathcal{M}}H_{3}(\mathcal{M}_{1},\mathcal{Z})-\nabla_{\mathcal{M}}H_{3}(\mathcal{M}_{2},\mathcal{Z})\|_{F}\nonumber
	\\&&=\|\mu(\mathcal{M}_{1}-\mathcal{Z}-\frac{\mathcal{Q}}{\mu})-\mu(\mathcal{M}_{2}-\mathcal{Z}-\frac{\mathcal{Q}}{\mu})\|_{F}\nonumber
	\\&&=\mu\|\mathcal{M}_{1}-\mathcal{M}_{2}\|_{F},\label{TCCM}
	\end{eqnarray}
	and for any fixed $\mathcal{M}$, we get that
	\begin{eqnarray}
	&&\|\nabla_{\mathcal{Z}}H_{3}(\mathcal{M},\mathcal{Z}_{1})-\nabla_{\mathcal{Z}}H_{3}(\mathcal{M},\mathcal{Z}_{2})\|_{F}\nonumber
	\\&&=\|\mu(\mathcal{Z}_{1}-\mathcal{M}+\frac{\mathcal{Q}}{\mu})-\mu(\mathcal{Z}_{2}-\mathcal{M}+\frac{\mathcal{Q}}{\mu})\|_{F}\nonumber
	\\&&=\mu\|(\mathcal{Z}_{1}-\mathcal{Z}_{2})\|_{F}.\label{TCCZ}
	\end{eqnarray}
	(\ref{TCCM}) and (\ref{TCCZ}) imply that the gradient of $H_{3}(\mathcal{M},\mathcal{Z})$ is Lipschitz continuous block-wise. Note that $H_{3}(\mathcal{M},\mathcal{Z})$ is twice continuously differentiable, which brings that $\nabla H_{3}(\mathcal{M},\mathcal{Z})$ is Lipschitz continuous on bounded subsets of $\mathbb{R}^{\mathit{I}_{1}\times\mathit{I}_{2}\times\cdots\times\mathit{I}_{N}}\times\mathbb{R}^{\mathit{I}_{1}\times\mathit{I}_{2}\times\cdots\times\mathit{I}_{N}}$ \cite{bolte2014proximal}. So
	\begin{eqnarray}
	&&H_{3}(\mathcal{M}^{k+1})+H_{1}(\mathcal{M}^{k+1})+\frac{(\gamma_{1}-1)\mu}{2}\|\mathcal{M}^{k+1}-\mathcal{M}^{k}\|_{F}^{2}\nonumber
	\\&&\leqslant H_{3}(\mathcal{M}^{k})+H_{1}(\mathcal{M}^{k}).
	\end{eqnarray}
	From \cite{bolte2014proximal}, we get that
	\begin{eqnarray}
	&&LAG(\mathcal{Z}^{k+1},\mathcal{M}^{k+1},W^{k+1},\bar{\Lambda}^{k+1})\nonumber
	\\&&+\frac{(\gamma_{1}-1)\mu}{2}\|\mathcal{M}^{k+1}-\mathcal{M}^{k}\|_{F}^{2}+\frac{\rho^{k}}{2}(\|W^{k+1}-W^{k}\|_{F}^{2})\nonumber
	\\&&+\frac{\rho^{k}}{2}(\|\bar{\Lambda}^{k+1}-\bar{\Lambda}^{k}\|_{F}^{2}+\|\mathcal{Z}^{k+1}-\mathcal{Z}^{k}\|_{F}^{2})\nonumber
	\\&&\leqslant LAG(\mathcal{Z}^{k},\mathcal{M}^{k},W^{k},\bar{\Lambda}^{k}),\label{TCLE}
	\end{eqnarray}
	and 
	\begin{eqnarray}
	&&\lim_{k\to+\infty}\|\mathcal{M}^{k+1}-\mathcal{M}^{k}\|_{F}\nonumber
	\\&&=\lim_{k\to+\infty}\|W^{k+1}-W^{k}\|_{F}\nonumber
	\\&&=\lim_{k\to+\infty}\|\bar{\Lambda}^{k+1}-\bar{\Lambda}^{k}\|_{F}\nonumber
	\\&&=\lim_{k\to+\infty}\|\mathcal{Z}^{k+1}-\mathcal{Z}^{k}\|_{F}=0.\label{COTC}
	\end{eqnarray}
\\ (a) Assume that there exists a subsequence $\{(\mathcal{Z}^{k_{j}},\mathcal{M}^{k_{j}},W^{k_{j}},\bar{\Lambda}^{k_{j}})\}$ of
$\{(\mathcal{Z}^{k},\mathcal{M}^{k},W^{k},\bar{\Lambda}^{k})\}$ such that $\{(\mathcal{Z}^{k_{j}},\mathcal{M}^{k_{j}},W^{k_{j}},\bar{\Lambda}^{k_{j}})\}$
converges to $(\mathcal{Z}^{\star},\mathcal{M}^{\star},W^{\star},\bar{\Lambda}^{\star})$ as $j\to+\infty$. By (\ref{COTC}), we have that $\{(\mathcal{Z}^{k_{j}},\mathcal{M}^{k_{j}},W^{k_{j}},\bar{\Lambda}^{k_{j}})\}$
also converges
to $(\mathcal{Z}^{\star},\mathcal{M}^{\star},W^{\star},\bar{\Lambda}^{\star})$ as $j\to+\infty$. Moreover, the optimality conditions of (\ref{PA}) gives that
\begin{eqnarray}
0=\nabla H_{1}(W^{k_{j}+1})+\nabla H_{2}(W^{k_{j}+1})+ \rho^{k}(W^{k_{j}+1}-W^{k_{j}}),\nonumber
\end{eqnarray}
\begin{eqnarray}
0\in\partial H_{1}(\mathcal{M}^{k_{j}+1})-\mu(\mathcal{Z}-\mathcal{M}^{k_{j}}+\frac{\mathcal{Q}}{\mu})+\rho_{1}^{k}(\mathcal{M}^{k_{j}+1}-\mathcal{M}^{k_{j}}),\nonumber
\end{eqnarray}
\begin{eqnarray}
0=\nabla H_{2}(\bar{\Lambda}^{k_{j}+1})+ \rho^{k}(\bar{\Lambda}^{k_{j}+1}-\bar{\Lambda}^{k_{j}}),\nonumber
\end{eqnarray}
\begin{eqnarray}
0=\nabla H_{3}(\mathcal{Z}^{k_{j}+1})+\rho^{k}(\mathcal{Z}^{k_{j}+1}-\mathcal{Z}^{k_{j}}),\nonumber
\end{eqnarray}
When $j\to+\infty$, by \cite{clarke1990optimization}, we get that
\begin{eqnarray}
0=\nabla H_{1}(W^{\star})+\nabla H_{2}(W^{\star}),\nonumber
\end{eqnarray}
\begin{eqnarray}
0\in\partial H_{1}(\mathcal{M}^{\star})-\mu(\mathcal{Z}-\mathcal{M}^{\star}+\frac{\mathcal{Q}}{\mu})=\partial H_{1}(\mathcal{M}^{\star})+\nabla H_{3}(\mathcal{M}^{\star}),\nonumber
\end{eqnarray}
\begin{eqnarray}
0=\nabla H_{2}(\bar{\Lambda}^{\star}),\nonumber
\\0=\nabla H_{3}(\mathcal{Z}^{\star}),\nonumber
\end{eqnarray}
Therefore, we obtain that
\begin{eqnarray}
(0,0,0,0)\in\partial LAG(\mathcal{Z}^{\star},\mathcal{M}^{\star},W^{\star},\bar{\Lambda}^{\star})
\end{eqnarray}
which implies that $(\mathcal{Z}^{\star},\mathcal{M}^{\star},W^{\star},\bar{\Lambda}^{\star})$ is a critial point of $LAG(\mathcal{Z},\mathcal{M},W,\bar{\Lambda})$.
\\ (b) By the definition of $LAG(\mathcal{Z},\mathcal{M},W,\bar{\Lambda})$, we have that $LAG(\mathcal{Z},\mathcal{M},W,\bar{\Lambda})\to+\infty$ as $\|(\mathcal{Z},\mathcal{M},W,\bar{\Lambda})\|_{F}\to+\infty$. Suppose that $(\mathcal{Z}^{k},\mathcal{M}^{k},W^{k},\bar{\Lambda}^{k})$ is unbounded, i.e., $\|(\mathcal{Z}^{k},\mathcal{M}^{k},W^{k},\bar{\Lambda}^{k})\|_{F}\to+\infty$, we derive that $LAG(\mathcal{Z}^{k},\mathcal{M}^{k},W^{k},\bar{\Lambda}^{k})\to+\infty$. However, it follows from (\ref{TCLE}) that $LAG(\mathcal{Z}^{k},\mathcal{M}^{k},W^{k},\bar{\Lambda}^{k})$ is upper bounded. Therefore, the
sequence $\{(\mathcal{Z}^{k},\mathcal{M}^{k},W^{k},\bar{\Lambda}^{k})\}$ is bounded. From \cite{ochs2015iteratively}, Logarithmic function is KL function. Thus, $H_{1}$ also KL function. Notice that $H_{2}$ and $H_{3}$ are KL function, we have that $LAG(\mathcal{Z}^{k},\mathcal{M}^{k},W^{k},\bar{\Lambda}^{k})$ is also a KL function \cite{bolte2014proximal}. Then by \cite{bolte2014proximal}, we obtain that the sequence $\{(\mathcal{Z}^{k},\mathcal{M}^{k},W^{k},\bar{\Lambda}^{k})\}$ converges to a critical point of (\ref{FTC1}).
\end{IEEEproof}
\begin{theorem} Suppose that $\rho_{1}=\gamma_{1}\mu$ with $\gamma_{1}>1$. Let the sequence $\{(\mathcal{L},\mathcal{G},W,\bar{\Lambda},\mathcal{E},\mathcal{N})\}$ be generated by Algorithm \ref{TPRCA}. Then,
	\\ (a) any accumulation point of the sequence $\{(\mathcal{L},\mathcal{G},W,\bar{\Lambda},\mathcal{E},\mathcal{N})\}$ is a critical point of (\ref{FPCA1}).
	\\ (b) if $H_{1}$ is KL functions and coercive, the sequence $\{(\mathcal{L},\mathcal{G},W,\bar{\Lambda},\mathcal{E},\mathcal{N})\}$ converges to a critical point of (\ref{FPCA1}).
\end{theorem}
\begin{IEEEproof}
	Compared with LRTC, the TRPCA algorithm has two more variables, $\mathcal{E}$ and $\mathcal{N}$, but its solutions are all convex functions. Therefore, the convergence proof of the TRPCA algorithm is similar to that of the LRTC algorithm, and will not be repeated here.
\end{IEEEproof}
\section{Experiments}
We evaluate the performance of the proposed EMLCP-based LRTC and TRPCA methods. All methods are tested on real-world data. We employ the peak signal-to-noise rate (PSNR) value, the structural similarity (SSIM) value \cite{1284395}, the feature similarity (FSIM) value \cite{5705575}, and erreur relative globale adimensionnelle de synth$\grave{e}$se (ERGAS) value \cite{2432002352} to measure the quality of the recovered results. The PSNR, SSIM and FSIM value are the bigger the better, and the ERGAS value is the smaller the better. For simplicity, EMLCP-based LRTC and EMLCP-based TRPCA are denoted as EMLCP. All tests are implemented on the Windows 10 platform and MATLAB (R2019a) with an Intel Core i7-10875H 2.30 GHz and 32 GB of RAM.

\subsection{Low-rank tensor completion}
In this section, we test three kinds of real-world data: MSI, MRI and CV. The method for sampling the data is purely random sampling. The comparative LRTC methods are as follows: HaLRTC \cite{6138863}, and LRTCTV-I \cite{3120171} represent state-of-the-art for the Tucker-decomposition-based methods; TNN \cite{7782758}, PSTNN \cite{1122020112680}, FTNN \cite{9115254}, WSTNN \cite{2020170}, and nonconvex WSTNN \cite{7342019749} represent state-of-the-art for the t-SVD-based methods; and minmax concave plus penalty-based TC method (McpTC) \cite{2612015273}. Since the TNN, PSTNN, and FTNN methods are only applicable to three-order tensors, in all four-order tensor tests, we first reshape the four-order tensor into three-order tensors and then test the performances of these methods. It is not difficult to find that the NWSTNN method in the comparison method adopts the non-convex relaxation of the Logarithmic function, and the results obtained by comparing with such method are consistent with our theory property.

\subsubsection{MSI completion}
\begin{figure*}[!h] 
	\centering  
	\vspace{0cm} 
	\subfloat[]{
		\begin{minipage}[b]{0.075\linewidth}
			\includegraphics[width=1\linewidth]{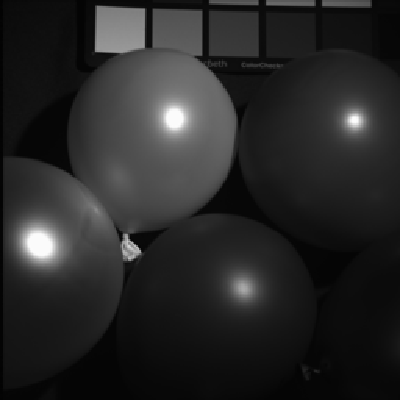}\\
			
			\includegraphics[width=1\linewidth]{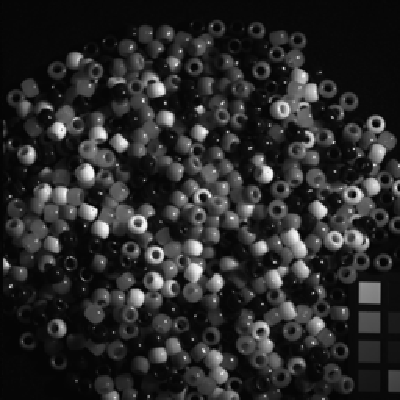}\\
			
			\includegraphics[width=1\linewidth]{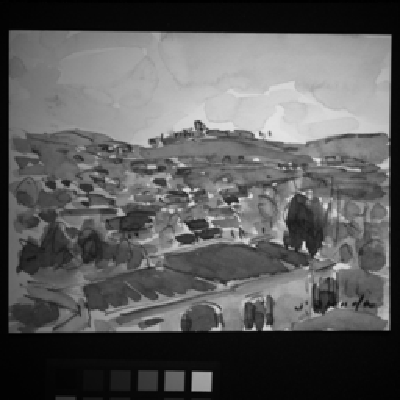}
	\end{minipage}}
	\subfloat[]{
		\begin{minipage}[b]{0.075\linewidth}
			\includegraphics[width=1\linewidth]{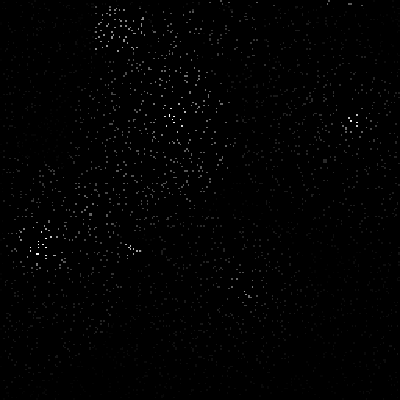}\\
			
			\includegraphics[width=1\linewidth]{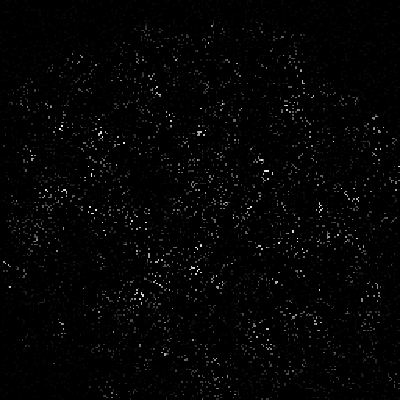}\\
			
			\includegraphics[width=1\linewidth]{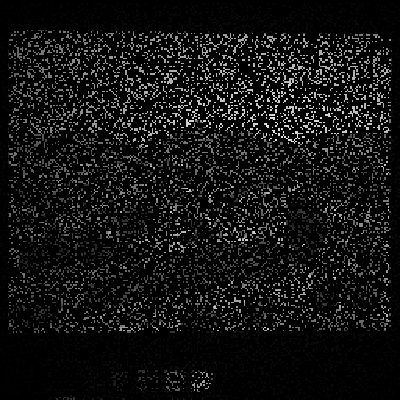}
	\end{minipage}}
\subfloat[]{
\begin{minipage}[b]{0.075\linewidth}
\includegraphics[width=1\linewidth]{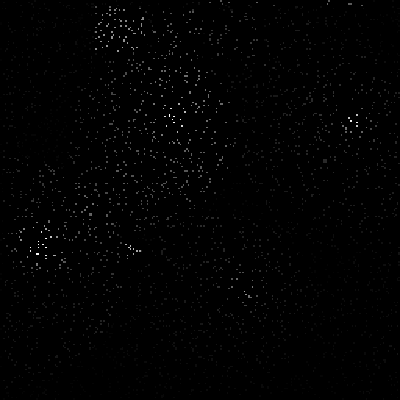}\\

\includegraphics[width=1\linewidth]{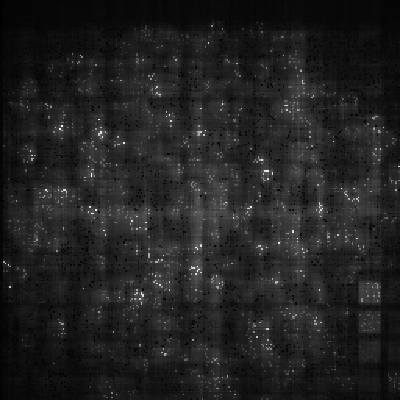}\\

\includegraphics[width=1\linewidth]{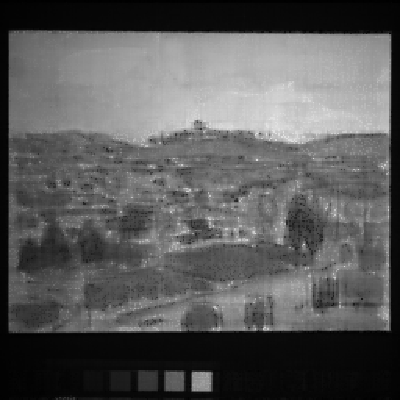}
\end{minipage}}
\subfloat[]{
\begin{minipage}[b]{0.075\linewidth}
\includegraphics[width=1\linewidth]{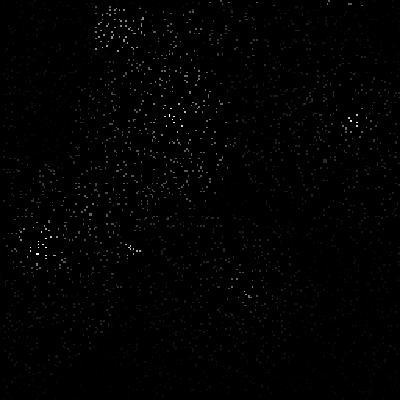}\\

\includegraphics[width=1\linewidth]{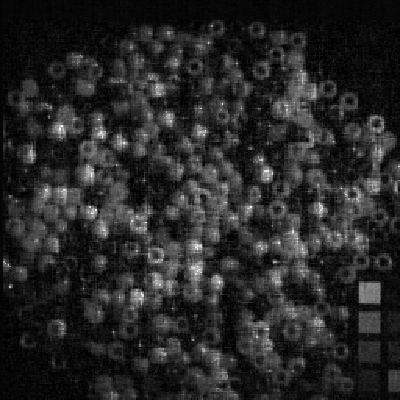}\\

\includegraphics[width=1\linewidth]{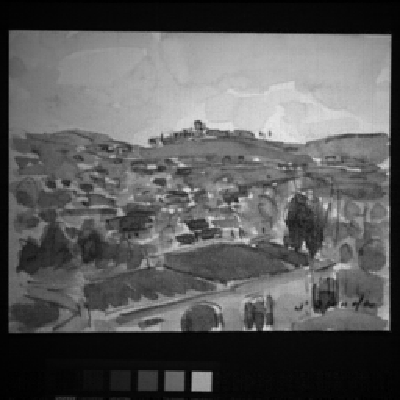}
\end{minipage}}
\subfloat[]{
\begin{minipage}[b]{0.075\linewidth}
\includegraphics[width=1\linewidth]{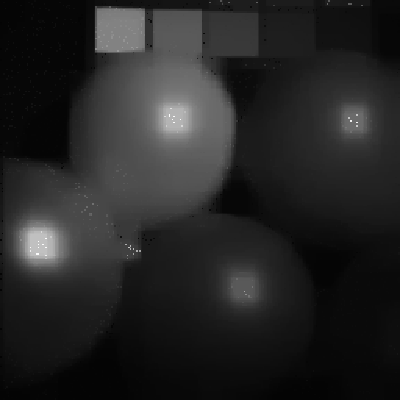}\\

\includegraphics[width=1\linewidth]{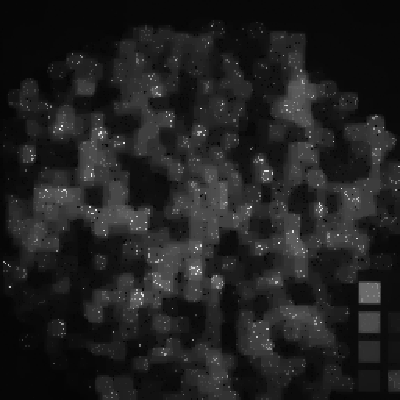}\\

\includegraphics[width=1\linewidth]{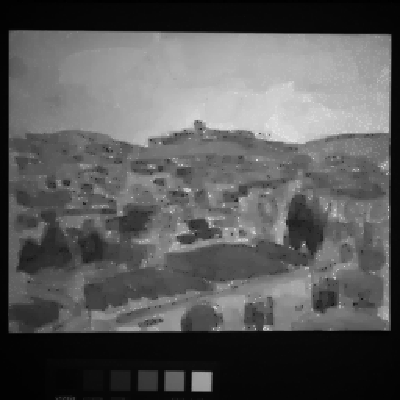}
\end{minipage}}
\subfloat[]{
\begin{minipage}[b]{0.075\linewidth}
\includegraphics[width=1\linewidth]{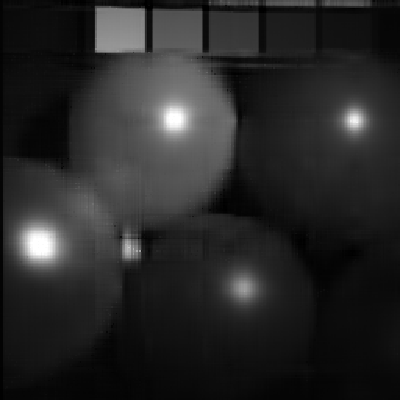}\\

\includegraphics[width=1\linewidth]{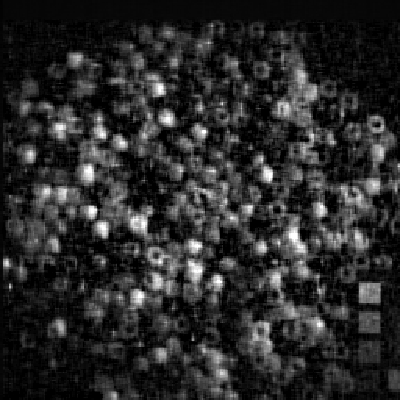}\\

\includegraphics[width=1\linewidth]{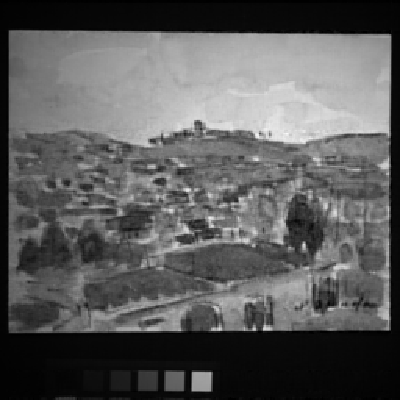}
\end{minipage}}
\subfloat[]{
\begin{minipage}[b]{0.075\linewidth}
\includegraphics[width=1\linewidth]{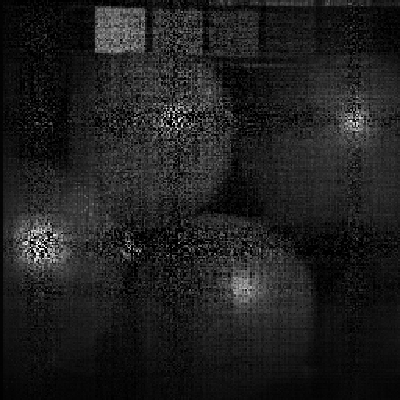}\\

\includegraphics[width=1\linewidth]{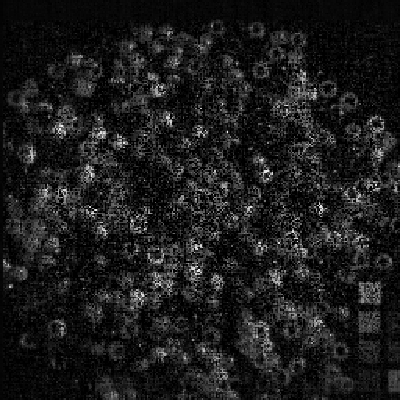}\\

\includegraphics[width=1\linewidth]{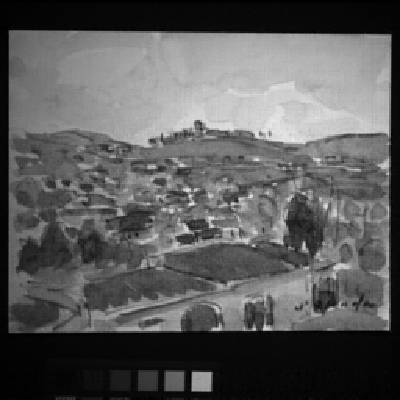}
\end{minipage}}
\subfloat[]{
\begin{minipage}[b]{0.075\linewidth}
\includegraphics[width=1\linewidth]{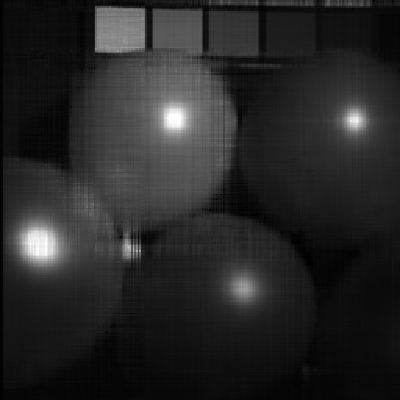}\\

\includegraphics[width=1\linewidth]{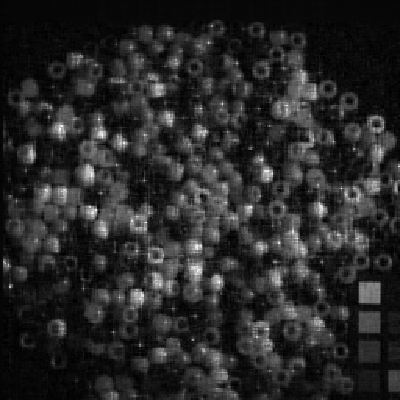}\\

\includegraphics[width=1\linewidth]{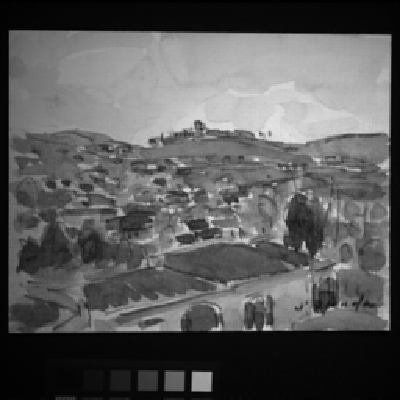}
\end{minipage}}
\subfloat[]{
\begin{minipage}[b]{0.075\linewidth}
\includegraphics[width=1\linewidth]{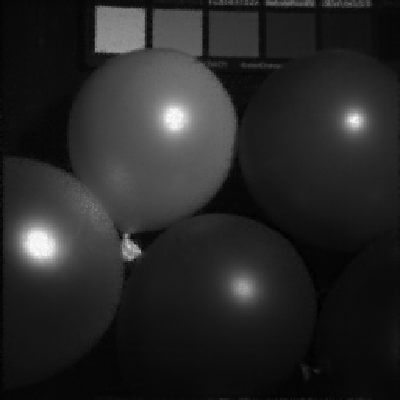}\\

\includegraphics[width=1\linewidth]{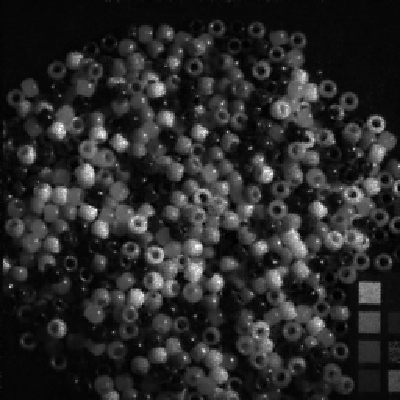}\\

\includegraphics[width=1\linewidth]{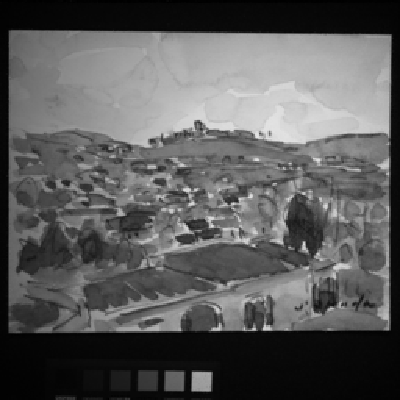}
\end{minipage}}
\subfloat[]{
\begin{minipage}[b]{0.075\linewidth}
\includegraphics[width=1\linewidth]{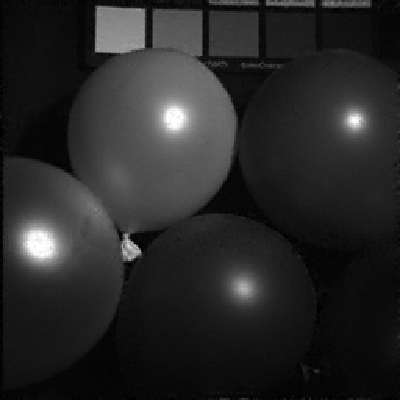}\\

\includegraphics[width=1\linewidth]{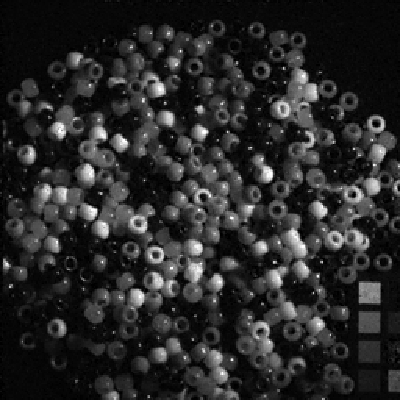}\\

\includegraphics[width=1\linewidth]{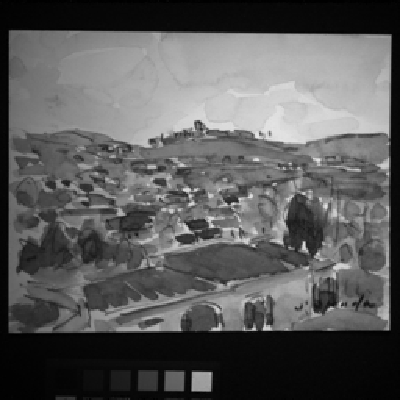}
\end{minipage}}
\subfloat[]{
\begin{minipage}[b]{0.075\linewidth}
\includegraphics[width=1\linewidth]{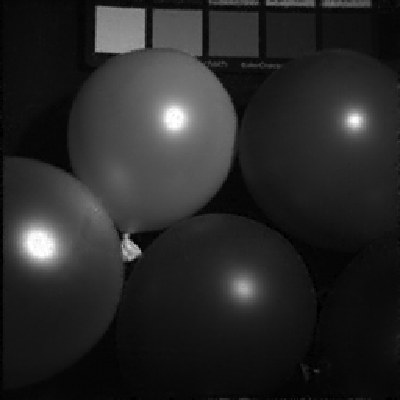}\\

\includegraphics[width=1\linewidth]{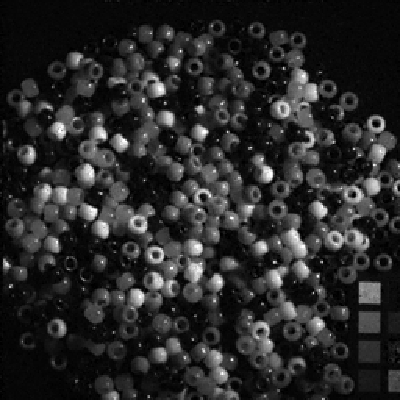}\\

\includegraphics[width=1\linewidth]{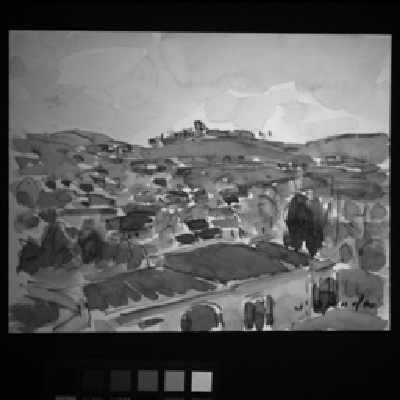}
\end{minipage}}
	\caption{(a) Original image. (b) Obeserved image. (c) HaLRTC. (d) TNN. (e) LRTCTV-I. (f) McpTC. (g) PSTNN. (h) FTNN. (i) WSTNN. (j) NWSTNN. (k) EMLCP. SR: top row is 5\%, middle row is 10\% and last row is 20\%. The rows of MSIs are in order: balloons, beads, watercolors. The corresponding bands in each row are: 31, 20, 10.}
	\label{MSITC}
\end{figure*}

We test 32 MSIs in the dataset CAVE\footnote{http://www.cs.columbia.edu/CAVE/databases/multispectral/}. All testing data are of size $256\times256\times31$. In Fig.\ref{MSITC}, we randomly select three from 32 MSIs, and brings the different sampling rate and different band visual results. The individual MSI names and their corresponding bands are written in the caption of Fig.\ref{MSITC}. As can be seen from Fig.\ref{MSITC}, the visual effect of the EMLCP method is superior to the NWSTNN method under all sample rate, which is consistent with our theory. To further highlight the superiority of our method, the average quantitative results of 32 MSIs are listed in Table \ref{MSITC1}. The results show that the PSNR value of our algorithms is 0.4dB higher than that of the suboptimal method when the sampling rate is 20\%, and even reaches 0.8dB when the sampling rate is 5\%. More experimental results are available in the Appendix K.
\begin{table*}[]
		\caption{The average PSNR, SSIM, FSIM and ERGAS values for 32 MSIs tested by observed and the nine utilized LRTC methods.}
	\resizebox{\textwidth}{!}{
	\begin{tabular}{|c|cccc|cccc|cccc|}
		\hline
		SR       & \multicolumn{4}{c|}{5\%}                                                                        & \multicolumn{4}{c|}{10\%}                                                                       & \multicolumn{4}{c|}{20\%}                                                                       \\ \hline
		Method   & \multicolumn{1}{c|}{PSNR}   & \multicolumn{1}{c|}{SSIM}  & \multicolumn{1}{c|}{FSIM}  & ERGAS   & \multicolumn{1}{c|}{PSNR}   & \multicolumn{1}{c|}{SSIM}  & \multicolumn{1}{c|}{FSIM}  & ERGAS   & \multicolumn{1}{c|}{PSNR}   & \multicolumn{1}{c|}{SSIM}  & \multicolumn{1}{c|}{FSIM}  & ERGAS   \\ \hline
		Observed & \multicolumn{1}{c|}{15.438} & \multicolumn{1}{c|}{0.153} & \multicolumn{1}{c|}{0.644} & 845.388 & \multicolumn{1}{c|}{15.673} & \multicolumn{1}{c|}{0.194} & \multicolumn{1}{c|}{0.646} & 822.788 & \multicolumn{1}{c|}{16.184} & \multicolumn{1}{c|}{0.269} & \multicolumn{1}{c|}{0.650} & 775.866 \\ \hline
		HaLRTC   & \multicolumn{1}{c|}{18.112} & \multicolumn{1}{c|}{0.285} & \multicolumn{1}{c|}{0.697} & 689.482 & \multicolumn{1}{c|}{22.694} & \multicolumn{1}{c|}{0.527} & \multicolumn{1}{c|}{0.786} & 478.325 & \multicolumn{1}{c|}{32.175} & \multicolumn{1}{c|}{0.835} & \multicolumn{1}{c|}{0.910} & 190.848 \\ \hline
		TNN      & \multicolumn{1}{c|}{17.986} & \multicolumn{1}{c|}{0.247} & \multicolumn{1}{c|}{0.685} & 726.893 & \multicolumn{1}{c|}{28.627} & \multicolumn{1}{c|}{0.678} & \multicolumn{1}{c|}{0.861} & 314.352 & \multicolumn{1}{c|}{40.170} & \multicolumn{1}{c|}{0.964} & \multicolumn{1}{c|}{0.972} & 59.018  \\ \hline
		LRTCTV-I & \multicolumn{1}{c|}{25.894} & \multicolumn{1}{c|}{0.800} & \multicolumn{1}{c|}{0.835} & 276.620 & \multicolumn{1}{c|}{30.709} & \multicolumn{1}{c|}{0.890} & \multicolumn{1}{c|}{0.906} & 162.567 & \multicolumn{1}{c|}{35.486} & \multicolumn{1}{c|}{0.949} & \multicolumn{1}{c|}{0.957} & 94.646  \\ \hline
		McpTC    & \multicolumn{1}{c|}{32.459} & \multicolumn{1}{c|}{0.875} & \multicolumn{1}{c|}{0.909} & 133.472 & \multicolumn{1}{c|}{35.959} & \multicolumn{1}{c|}{0.925} & \multicolumn{1}{c|}{0.943} & 91.788  & \multicolumn{1}{c|}{40.518} & \multicolumn{1}{c|}{0.964} & \multicolumn{1}{c|}{0.972} & 56.083  \\ \hline
		PSTNN    & \multicolumn{1}{c|}{18.713} & \multicolumn{1}{c|}{0.474} & \multicolumn{1}{c|}{0.650} & 574.637 & \multicolumn{1}{c|}{23.239} & \multicolumn{1}{c|}{0.683} & \multicolumn{1}{c|}{0.783} & 352.012 & \multicolumn{1}{c|}{34.206} & \multicolumn{1}{c|}{0.924} & \multicolumn{1}{c|}{0.942} & 117.472 \\ \hline
		FTNN     & \multicolumn{1}{c|}{32.620} & \multicolumn{1}{c|}{0.899} & \multicolumn{1}{c|}{0.924} & 131.871 & \multicolumn{1}{c|}{37.182} & \multicolumn{1}{c|}{0.954} & \multicolumn{1}{c|}{0.963} & 78.694  & \multicolumn{1}{c|}{43.002} & \multicolumn{1}{c|}{0.984} & \multicolumn{1}{c|}{0.987} & 41.625  \\ \hline
		WSTNN    & \multicolumn{1}{c|}{31.439} & \multicolumn{1}{c|}{0.806} & \multicolumn{1}{c|}{0.911} & 208.988 & \multicolumn{1}{c|}{40.170} & \multicolumn{1}{c|}{0.981} & \multicolumn{1}{c|}{0.981} & 52.895  & \multicolumn{1}{c|}{47.059} & \multicolumn{1}{c|}{0.995} & \multicolumn{1}{c|}{0.995} & 24.914  \\ \hline
		NWSTNN   & \multicolumn{1}{c|}{37.417} & \multicolumn{1}{c|}{0.945} & \multicolumn{1}{c|}{0.950} & 71.261  & \multicolumn{1}{c|}{43.704} & \multicolumn{1}{c|}{0.985} & \multicolumn{1}{c|}{0.985} & 35.779  & \multicolumn{1}{c|}{51.362} & \multicolumn{1}{c|}{0.997} & \multicolumn{1}{c|}{0.997} & 15.572  \\ \hline
		EMLCP    & \multicolumn{1}{c|}{38.298} & \multicolumn{1}{c|}{0.962} & \multicolumn{1}{c|}{0.964} & 64.689  & \multicolumn{1}{c|}{44.340} & \multicolumn{1}{c|}{0.988} & \multicolumn{1}{c|}{0.988} & 33.329  & \multicolumn{1}{c|}{51.742} & \multicolumn{1}{c|}{0.997} & \multicolumn{1}{c|}{0.997} & 14.779  \\ \hline
	\end{tabular}}\label{MSITC1}
\end{table*}
\subsubsection{MRI completion}
\begin{figure*}[!h] 
	\centering  
	\vspace{0cm} 
	\subfloat[]{
		\begin{minipage}[b]{0.075\linewidth}
			\includegraphics[width=1\linewidth]{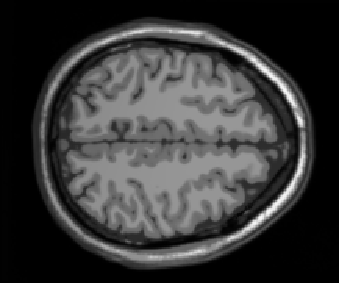}\\
			
			\includegraphics[width=1\linewidth]{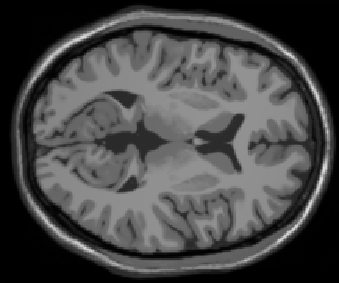}\\
			
			\includegraphics[width=1\linewidth]{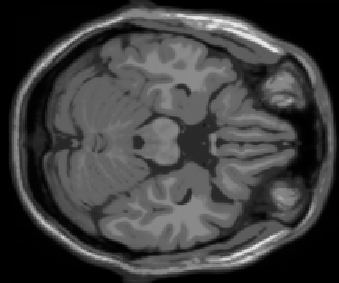}
	\end{minipage}}
	\subfloat[]{
		\begin{minipage}[b]{0.075\linewidth}
			\includegraphics[width=1\linewidth]{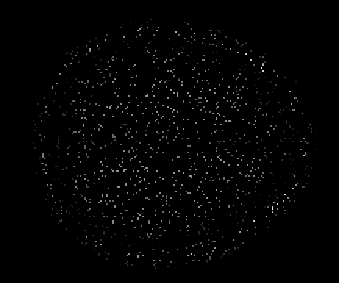}\\
			
			\includegraphics[width=1\linewidth]{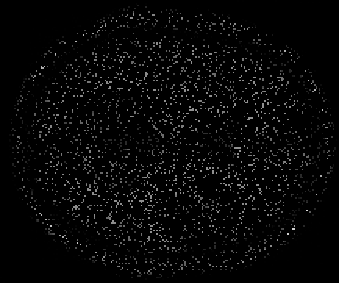}\\
			
			\includegraphics[width=1\linewidth]{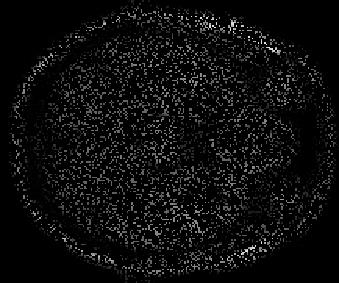}
	\end{minipage}}
\subfloat[]{
	\begin{minipage}[b]{0.075\linewidth}
		\includegraphics[width=1\linewidth]{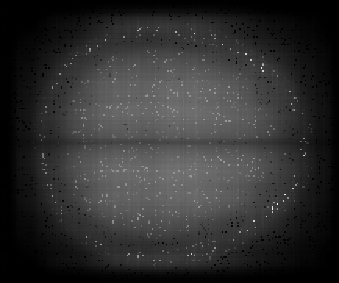}\\
		
		\includegraphics[width=1\linewidth]{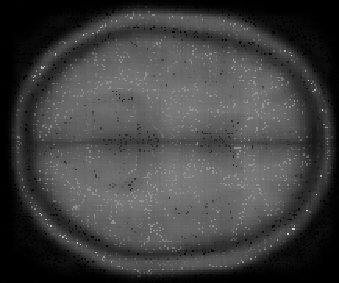}\\
		
		\includegraphics[width=1\linewidth]{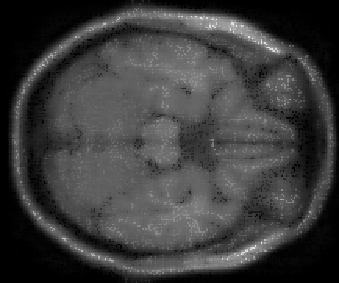}
\end{minipage}}
\subfloat[]{
	\begin{minipage}[b]{0.075\linewidth}
		\includegraphics[width=1\linewidth]{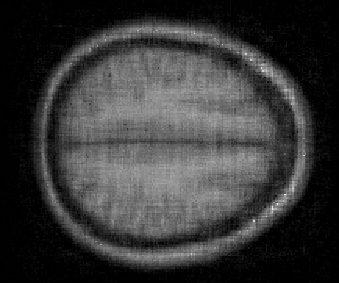}\\
		
		\includegraphics[width=1\linewidth]{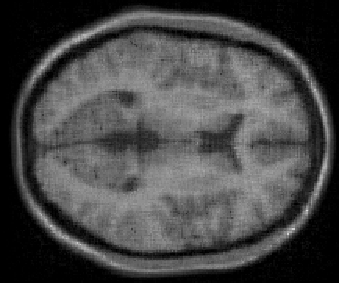}\\
		
		\includegraphics[width=1\linewidth]{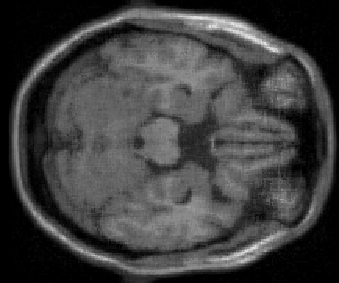}
\end{minipage}}
\subfloat[]{
	\begin{minipage}[b]{0.075\linewidth}
		\includegraphics[width=1\linewidth]{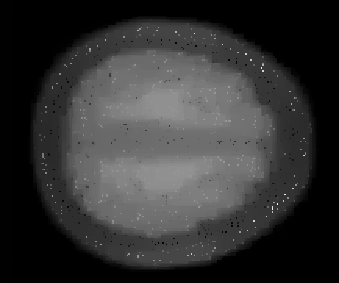}\\
		
		\includegraphics[width=1\linewidth]{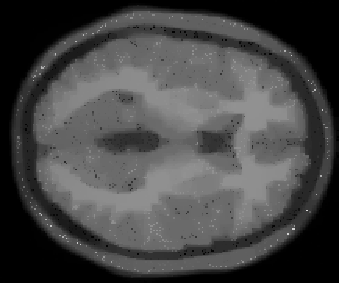}\\
		
		\includegraphics[width=1\linewidth]{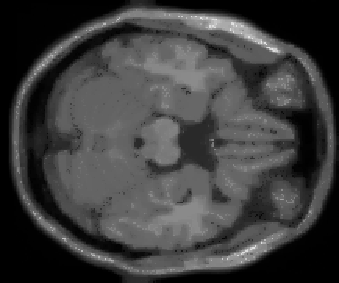}
\end{minipage}}
\subfloat[]{
	\begin{minipage}[b]{0.075\linewidth}
		\includegraphics[width=1\linewidth]{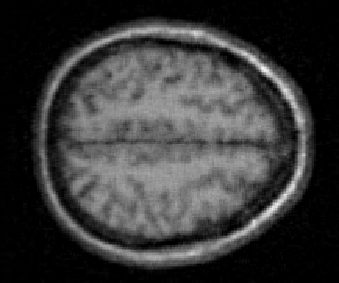}\\
		
		\includegraphics[width=1\linewidth]{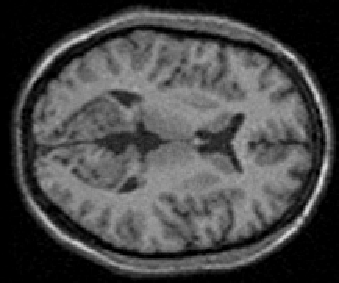}\\
		
		\includegraphics[width=1\linewidth]{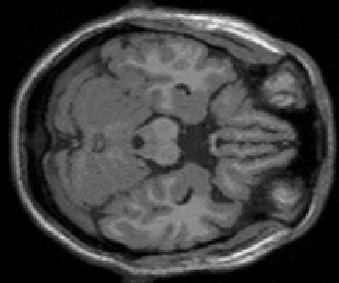}
\end{minipage}}
\subfloat[]{
	\begin{minipage}[b]{0.075\linewidth}
		\includegraphics[width=1\linewidth]{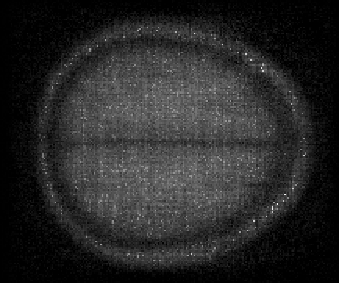}\\
		
		\includegraphics[width=1\linewidth]{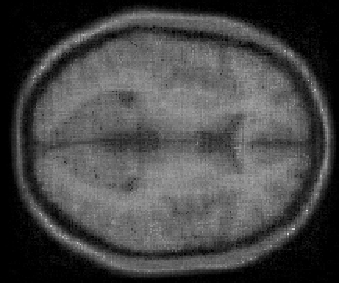}\\
		
		\includegraphics[width=1\linewidth]{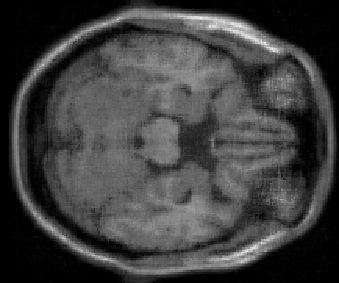}
\end{minipage}}
\subfloat[]{
	\begin{minipage}[b]{0.075\linewidth}
		\includegraphics[width=1\linewidth]{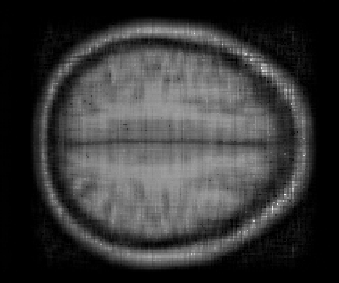}\\
		
		\includegraphics[width=1\linewidth]{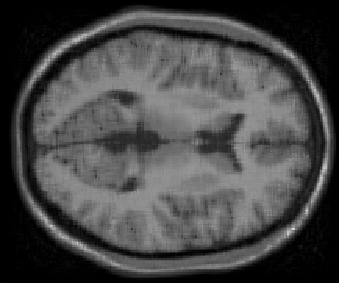}\\
		
		\includegraphics[width=1\linewidth]{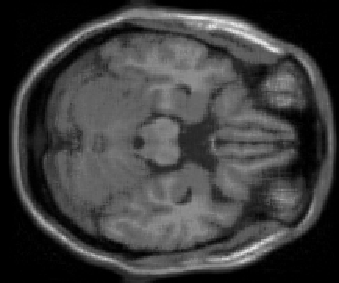}
\end{minipage}}
\subfloat[]{
	\begin{minipage}[b]{0.075\linewidth}
		\includegraphics[width=1\linewidth]{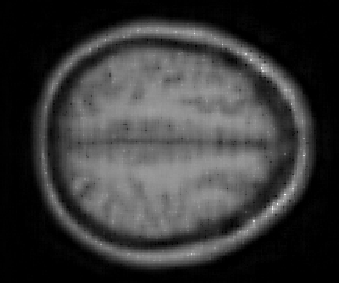}\\
		
		\includegraphics[width=1\linewidth]{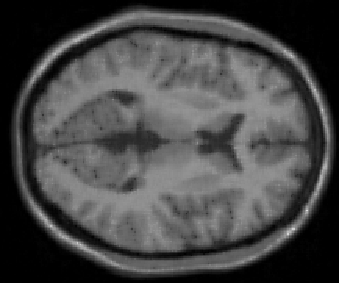}\\
		
		\includegraphics[width=1\linewidth]{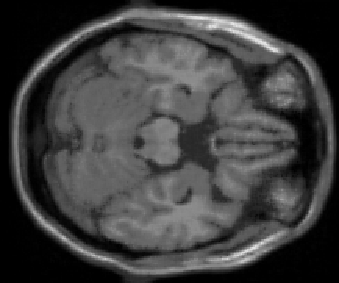}
\end{minipage}}
\subfloat[]{
	\begin{minipage}[b]{0.075\linewidth}
		\includegraphics[width=1\linewidth]{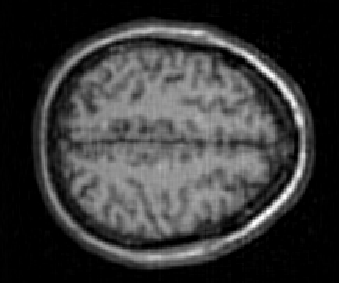}\\
		
		\includegraphics[width=1\linewidth]{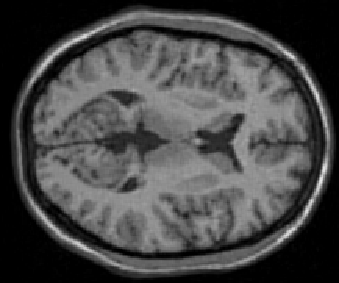}\\
		
		\includegraphics[width=1\linewidth]{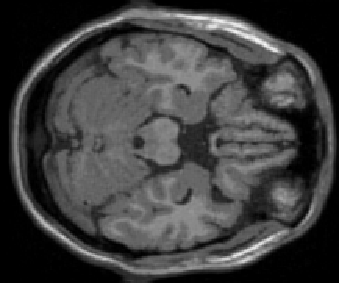}
\end{minipage}}
\subfloat[]{
	\begin{minipage}[b]{0.075\linewidth}
		\includegraphics[width=1\linewidth]{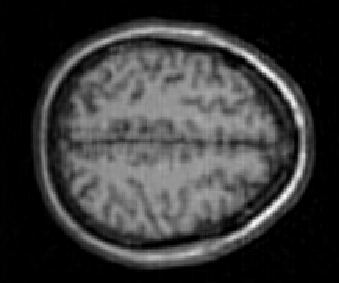}\\
		
		\includegraphics[width=1\linewidth]{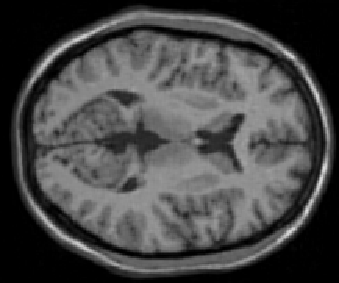}\\
		
		\includegraphics[width=1\linewidth]{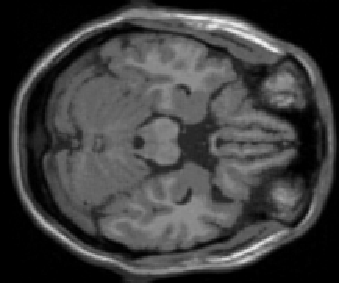}
\end{minipage}}
	\caption{(a) Original image. (b) Obeserved image. (c) HaLRTC. (d) TNN. (e) LRTCTV-I. (f) McpTC. (g) PSTNN. (h) FTNN. (i) WSTNN. (j) NWSTNN. (k) EMLCP. Each type of slice: the first row is the 120th slice with a sampling rate of 5\%, the second row is the 80th slice with a sampling rate of 10\%, and the third row is the 50th slice with a sampling rate of 20\%.}
	\label{MRITC}
\end{figure*}

We test the performance of the proposed method and the comparative method on MRI\footnote{http://brainweb.bic.mni.mcgill.ca/brainweb/selection\_normal.html} data with the size of $181\times217\times181$. First, we demonstrate the visual effect recovered by MRI data at sampling rates of 5\%, 10\% and 20\% in Fig.\ref{MRITC}. Our method is clearly superior to the comparative methods. Then, we list the average quantitative results of frontal slices of MRI restored by all methods at different sampling rates in Table \ref{MRITC1}. Obviously, the PSNR value of our method is at average 0.3dB higher than that of the suboptimal method, and the values of SSIM, FSIM and ERGAS are significantly better than that of the suboptimal method. 
\begin{table*}[]
		\caption{The PSNR, SSIM, FSIM and ERGAS values output by by observed and the nine utilized LRTC methods for MRI.}
	\resizebox{\textwidth}{!}{
	\begin{tabular}{|c|cccc|cccc|cccc|}
		\hline
		SR       & \multicolumn{4}{c|}{5\%}                                                                         & \multicolumn{4}{c|}{10\%}                                                                       & \multicolumn{4}{c|}{20\%}                                                                       \\ \hline
		Method   & \multicolumn{1}{c|}{PSNR}   & \multicolumn{1}{c|}{SSIM}  & \multicolumn{1}{c|}{FSIM}  & ERGAS    & \multicolumn{1}{c|}{PSNR}   & \multicolumn{1}{c|}{SSIM}  & \multicolumn{1}{c|}{FSIM}  & ERGAS   & \multicolumn{1}{c|}{PSNR}   & \multicolumn{1}{c|}{SSIM}  & \multicolumn{1}{c|}{FSIM}  & ERGAS   \\ \hline
		Observed & \multicolumn{1}{c|}{11.399} & \multicolumn{1}{c|}{0.310} & \multicolumn{1}{c|}{0.530} & 1021.071 & \multicolumn{1}{c|}{11.633} & \multicolumn{1}{c|}{0.323} & \multicolumn{1}{c|}{0.565} & 994.049 & \multicolumn{1}{c|}{12.149} & \multicolumn{1}{c|}{0.350} & \multicolumn{1}{c|}{0.613} & 936.747 \\ \hline
		HaLRTC   & \multicolumn{1}{c|}{17.372} & \multicolumn{1}{c|}{0.301} & \multicolumn{1}{c|}{0.638} & 532.927  & \multicolumn{1}{c|}{20.105} & \multicolumn{1}{c|}{0.439} & \multicolumn{1}{c|}{0.726} & 391.945 & \multicolumn{1}{c|}{24.451} & \multicolumn{1}{c|}{0.659} & \multicolumn{1}{c|}{0.829} & 235.019 \\ \hline
		TNN      & \multicolumn{1}{c|}{22.681} & \multicolumn{1}{c|}{0.470} & \multicolumn{1}{c|}{0.742} & 303.284  & \multicolumn{1}{c|}{26.064} & \multicolumn{1}{c|}{0.643} & \multicolumn{1}{c|}{0.812} & 205.410 & \multicolumn{1}{c|}{29.972} & \multicolumn{1}{c|}{0.798} & \multicolumn{1}{c|}{0.882} & 130.791 \\ \hline
		LRTCTV-I & \multicolumn{1}{c|}{19.400} & \multicolumn{1}{c|}{0.598} & \multicolumn{1}{c|}{0.702} & 431.241  & \multicolumn{1}{c|}{22.864} & \multicolumn{1}{c|}{0.749} & \multicolumn{1}{c|}{0.805} & 294.937 & \multicolumn{1}{c|}{28.236} & \multicolumn{1}{c|}{0.891} & \multicolumn{1}{c|}{0.908} & 155.272 \\ \hline
		McpTC    & \multicolumn{1}{c|}{27.931} & \multicolumn{1}{c|}{0.748} & \multicolumn{1}{c|}{0.843} & 154.029  & \multicolumn{1}{c|}{31.439} & \multicolumn{1}{c|}{0.844} & \multicolumn{1}{c|}{0.888} & 102.744 & \multicolumn{1}{c|}{35.576} & \multicolumn{1}{c|}{0.937} & \multicolumn{1}{c|}{0.941} & 63.906  \\ \hline
		PSTNN    & \multicolumn{1}{c|}{17.064} & \multicolumn{1}{c|}{0.243} & \multicolumn{1}{c|}{0.639} & 542.819  & \multicolumn{1}{c|}{22.870} & \multicolumn{1}{c|}{0.487} & \multicolumn{1}{c|}{0.757} & 297.337 & \multicolumn{1}{c|}{29.083} & \multicolumn{1}{c|}{0.772} & \multicolumn{1}{c|}{0.870} & 145.165 \\ \hline
		FTNN     & \multicolumn{1}{c|}{24.673} & \multicolumn{1}{c|}{0.687} & \multicolumn{1}{c|}{0.836} & 234.329  & \multicolumn{1}{c|}{28.297} & \multicolumn{1}{c|}{0.820} & \multicolumn{1}{c|}{0.896} & 152.733 & \multicolumn{1}{c|}{32.767} & \multicolumn{1}{c|}{0.919} & \multicolumn{1}{c|}{0.947} & 89.543  \\ \hline
		WSTNN    & \multicolumn{1}{c|}{25.524} & \multicolumn{1}{c|}{0.708} & \multicolumn{1}{c|}{0.825} & 211.315  & \multicolumn{1}{c|}{29.059} & \multicolumn{1}{c|}{0.837} & \multicolumn{1}{c|}{0.888} & 139.177 & \multicolumn{1}{c|}{33.497} & \multicolumn{1}{c|}{0.928} & \multicolumn{1}{c|}{0.940} & 82.851  \\ \hline
		NWSTNN   & \multicolumn{1}{c|}{30.222} & \multicolumn{1}{c|}{0.826} & \multicolumn{1}{c|}{0.884} & 119.820  & \multicolumn{1}{c|}{33.293} & \multicolumn{1}{c|}{0.902} & \multicolumn{1}{c|}{0.924} & 83.608  & \multicolumn{1}{c|}{36.860} & \multicolumn{1}{c|}{0.950} & \multicolumn{1}{c|}{0.956} & 54.962  \\ \hline
		EMLCP    & \multicolumn{1}{c|}{30.563} & \multicolumn{1}{c|}{0.850} & \multicolumn{1}{c|}{0.893} & 115.395  & \multicolumn{1}{c|}{33.643} & \multicolumn{1}{c|}{0.918} & \multicolumn{1}{c|}{0.932} & 80.590  & \multicolumn{1}{c|}{37.180} & \multicolumn{1}{c|}{0.959} & \multicolumn{1}{c|}{0.962} & 53.344  \\ \hline
	\end{tabular}}\label{MRITC1}
\end{table*}
\subsubsection{CV completion}
\begin{figure*}[!h] 
	\centering  
	\vspace{0cm} 
	\subfloat[]{
		\begin{minipage}[b]{0.075\linewidth}
			\includegraphics[width=1\linewidth]{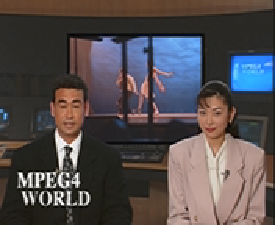}\\
			
			\includegraphics[width=1\linewidth]{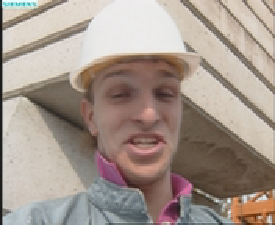}\\
			
			\includegraphics[width=1\linewidth]{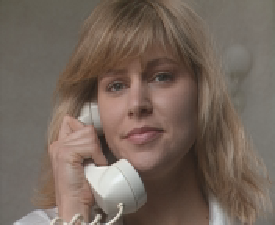}
	\end{minipage}}
	\subfloat[]{
		\begin{minipage}[b]{0.075\linewidth}
			\includegraphics[width=1\linewidth]{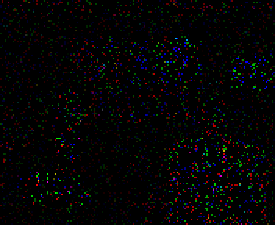}\\
			
			\includegraphics[width=1\linewidth]{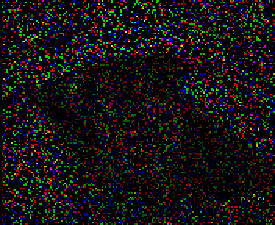}\\
			
			\includegraphics[width=1\linewidth]{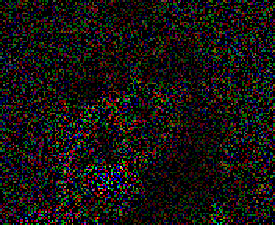}
	\end{minipage}}
	\subfloat[]{
		\begin{minipage}[b]{0.075\linewidth}
			\includegraphics[width=1\linewidth]{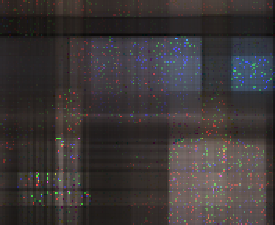}\\
			
			\includegraphics[width=1\linewidth]{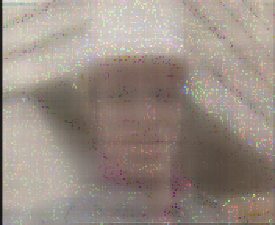}\\
			
			\includegraphics[width=1\linewidth]{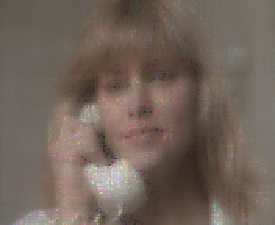}
	\end{minipage}}
\subfloat[]{
	\begin{minipage}[b]{0.075\linewidth}
		\includegraphics[width=1\linewidth]{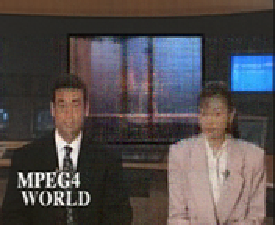}\\
		
		\includegraphics[width=1\linewidth]{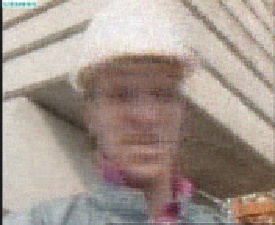}\\
		
		\includegraphics[width=1\linewidth]{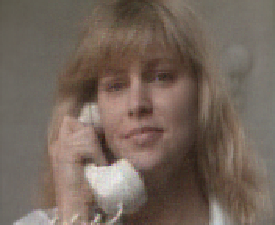}
\end{minipage}}
\subfloat[]{
	\begin{minipage}[b]{0.075\linewidth}
		\includegraphics[width=1\linewidth]{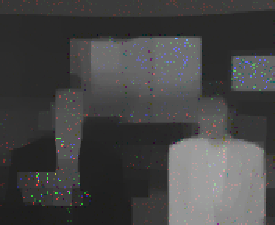}\\
		
		\includegraphics[width=1\linewidth]{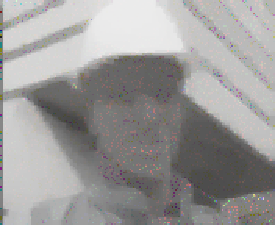}\\
		
		\includegraphics[width=1\linewidth]{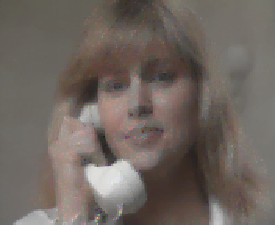}
\end{minipage}}
\subfloat[]{
	\begin{minipage}[b]{0.075\linewidth}
		\includegraphics[width=1\linewidth]{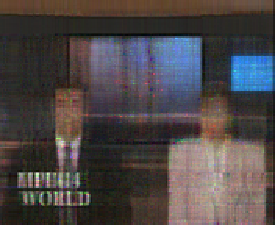}\\
		
		\includegraphics[width=1\linewidth]{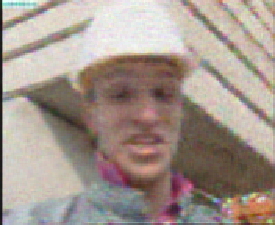}\\
		
		\includegraphics[width=1\linewidth]{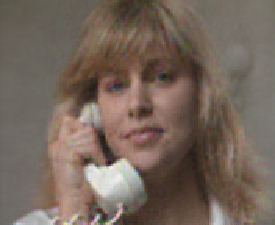}
\end{minipage}}
\subfloat[]{
	\begin{minipage}[b]{0.075\linewidth}
		\includegraphics[width=1\linewidth]{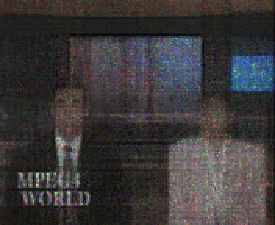}\\
		
		\includegraphics[width=1\linewidth]{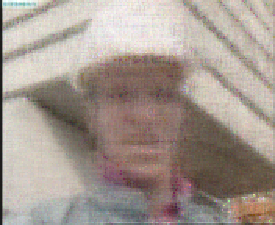}\\
		
		\includegraphics[width=1\linewidth]{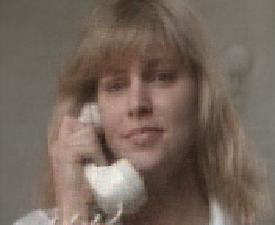}
\end{minipage}}
\subfloat[]{
	\begin{minipage}[b]{0.075\linewidth}
		\includegraphics[width=1\linewidth]{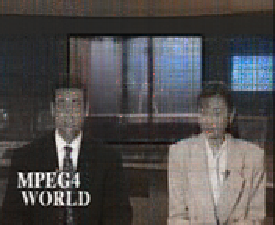}\\
		
		\includegraphics[width=1\linewidth]{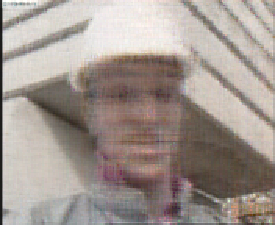}\\
		
		\includegraphics[width=1\linewidth]{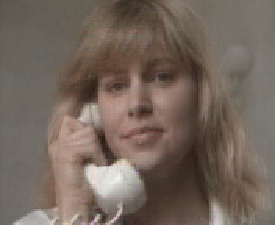}
\end{minipage}}
\subfloat[]{
	\begin{minipage}[b]{0.075\linewidth}
		\includegraphics[width=1\linewidth]{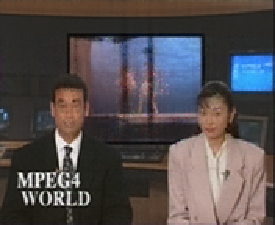}\\
		
		\includegraphics[width=1\linewidth]{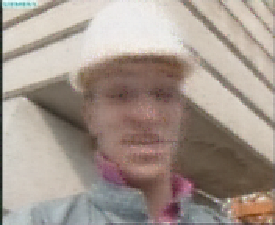}\\
		
		\includegraphics[width=1\linewidth]{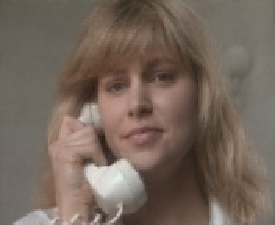}
\end{minipage}}
\subfloat[]{
	\begin{minipage}[b]{0.075\linewidth}
		\includegraphics[width=1\linewidth]{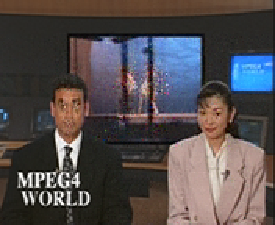}\\
		
		\includegraphics[width=1\linewidth]{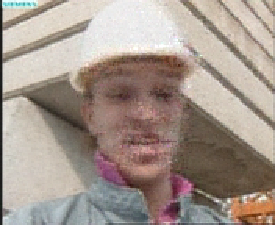}\\
		
		\includegraphics[width=1\linewidth]{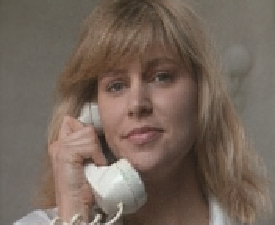}
\end{minipage}}
\subfloat[]{
	\begin{minipage}[b]{0.075\linewidth}
		\includegraphics[width=1\linewidth]{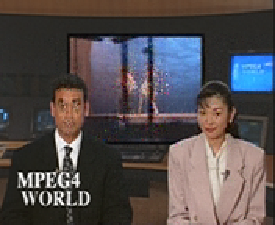}\\
		
		\includegraphics[width=1\linewidth]{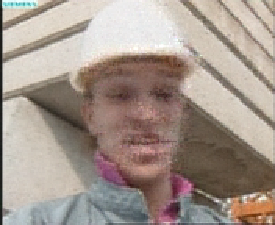}\\
		
		\includegraphics[width=1\linewidth]{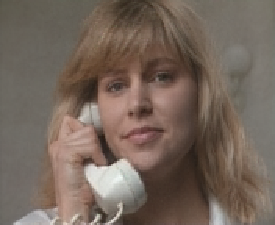}
\end{minipage}}
	\caption{(a) Original image. (b) Obeserved image. (c) HaLRTC. (d) TNN. (e) LRTCTV-I. (f) McpTC. (g) PSTNN. (h) FTNN. (i) WSTNN. (j) NWSTNN. (k) EMLCP. SR: top row is 5\%, middle row is 10\% and last row is 20\%. The rows of CVs are in order: the 15th frame of news, the 30th frame of foreman, the 45th frame of suzie.}
	\label{CVTC}
\end{figure*}

We test nine CVs\footnote{http://trace.eas.asu.edu/yuv/}(respectively named news, akiyo, hall, highway, foreman, container, coastguard, suzie, carphone) of size $144 \times 176 \times 3 \times 50$. Firstly, we list the average quantitative results of 9 CVs in Table \ref{CVTC1}. At this time, the suboptimal method is the NWSTNN method. The PSNR value of our method is average 0.4dB higher than it at three sampling rates. Furthermore, we demonstrate the visual results of 9 CVs in our experiment in Fig.\ref{CVTC}, in which the number of frames and sampling rate corresponding to each CV are described in the caption of Fig.\ref{CVTC}. It is not hard to see from the picture that the recovery of our method on the vision effect is better. More experimental results are available in the Appendix L.
\begin{table*}[]
	\caption{The average PSNR, SSIM, FSIM and ERGAS values for 9 CVs tested by observed and the nine utilized LRTC methods.}
	\resizebox{\textwidth}{!}{
	\begin{tabular}{|c|cccc|cccc|cccc|}
		\hline
		SR       & \multicolumn{4}{c|}{5\%}                                                                         & \multicolumn{4}{c|}{10\%}                                                                        & \multicolumn{4}{c|}{20\%}                                                                        \\ \hline
		Method   & \multicolumn{1}{c|}{PSNR}   & \multicolumn{1}{c|}{SSIM}  & \multicolumn{1}{c|}{FSIM}  & ERGAS    & \multicolumn{1}{c|}{PSNR}   & \multicolumn{1}{c|}{SSIM}  & \multicolumn{1}{c|}{FSIM}  & ERGAS    & \multicolumn{1}{c|}{PSNR}   & \multicolumn{1}{c|}{SSIM}  & \multicolumn{1}{c|}{FSIM}  & ERGAS    \\ \hline
		Observed & \multicolumn{1}{c|}{6.129}  & \multicolumn{1}{c|}{0.012} & \multicolumn{1}{c|}{0.431} & 1170.276 & \multicolumn{1}{c|}{6.363}  & \multicolumn{1}{c|}{0.019} & \multicolumn{1}{c|}{0.428} & 1139.091 & \multicolumn{1}{c|}{6.875}  & \multicolumn{1}{c|}{0.033} & \multicolumn{1}{c|}{0.426} & 1073.940 \\ \hline
		HaLRTC   & \multicolumn{1}{c|}{17.439} & \multicolumn{1}{c|}{0.497} & \multicolumn{1}{c|}{0.700} & 329.176  & \multicolumn{1}{c|}{21.207} & \multicolumn{1}{c|}{0.625} & \multicolumn{1}{c|}{0.776} & 214.604  & \multicolumn{1}{c|}{25.178} & \multicolumn{1}{c|}{0.775} & \multicolumn{1}{c|}{0.864} & 135.612  \\ \hline
		TNN      & \multicolumn{1}{c|}{26.940} & \multicolumn{1}{c|}{0.764} & \multicolumn{1}{c|}{0.881} & 114.770  & \multicolumn{1}{c|}{30.092} & \multicolumn{1}{c|}{0.845} & \multicolumn{1}{c|}{0.922} & 82.293   & \multicolumn{1}{c|}{33.193} & \multicolumn{1}{c|}{0.902} & \multicolumn{1}{c|}{0.950} & 59.164   \\ \hline
		LRTCTV-I & \multicolumn{1}{c|}{19.945} & \multicolumn{1}{c|}{0.598} & \multicolumn{1}{c|}{0.708} & 259.639  & \multicolumn{1}{c|}{21.864} & \multicolumn{1}{c|}{0.674} & \multicolumn{1}{c|}{0.786} & 213.126  & \multicolumn{1}{c|}{26.458} & \multicolumn{1}{c|}{0.826} & \multicolumn{1}{c|}{0.888} & 119.600  \\ \hline
		McpTC    & \multicolumn{1}{c|}{23.799} & \multicolumn{1}{c|}{0.669} & \multicolumn{1}{c|}{0.822} & 161.726  & \multicolumn{1}{c|}{28.480} & \multicolumn{1}{c|}{0.817} & \multicolumn{1}{c|}{0.898} & 93.541   & \multicolumn{1}{c|}{31.195} & \multicolumn{1}{c|}{0.885} & \multicolumn{1}{c|}{0.934} & 68.258   \\ \hline
		PSTNN    & \multicolumn{1}{c|}{15.274} & \multicolumn{1}{c|}{0.307} & \multicolumn{1}{c|}{0.670} & 409.255  & \multicolumn{1}{c|}{26.822} & \multicolumn{1}{c|}{0.776} & \multicolumn{1}{c|}{0.886} & 114.335  & \multicolumn{1}{c|}{32.739} & \multicolumn{1}{c|}{0.900} & \multicolumn{1}{c|}{0.948} & 61.799   \\ \hline
		FTNN     & \multicolumn{1}{c|}{25.563} & \multicolumn{1}{c|}{0.768} & \multicolumn{1}{c|}{0.872} & 133.678  & \multicolumn{1}{c|}{28.718} & \multicolumn{1}{c|}{0.856} & \multicolumn{1}{c|}{0.917} & 92.039   & \multicolumn{1}{c|}{32.209} & \multicolumn{1}{c|}{0.922} & \multicolumn{1}{c|}{0.952} & 61.699   \\ \hline
		WSTNN    & \multicolumn{1}{c|}{29.128} & \multicolumn{1}{c|}{0.869} & \multicolumn{1}{c|}{0.916} & 89.451   & \multicolumn{1}{c|}{32.341} & \multicolumn{1}{c|}{0.919} & \multicolumn{1}{c|}{0.948} & 63.735   & \multicolumn{1}{c|}{36.049} & \multicolumn{1}{c|}{0.957} & \multicolumn{1}{c|}{0.972} & 42.622   \\ \hline
		NWSTNN   & \multicolumn{1}{c|}{30.230} & \multicolumn{1}{c|}{0.845} & \multicolumn{1}{c|}{0.927} & 81.396   & \multicolumn{1}{c|}{34.002} & \multicolumn{1}{c|}{0.911} & \multicolumn{1}{c|}{0.956} & 54.680   & \multicolumn{1}{c|}{38.520} & \multicolumn{1}{c|}{0.960} & \multicolumn{1}{c|}{0.980} & 32.930   \\ \hline
		EMLCP    & \multicolumn{1}{c|}{30.872} & \multicolumn{1}{c|}{0.871} & \multicolumn{1}{c|}{0.934} & 75.578   & \multicolumn{1}{c|}{34.570} & \multicolumn{1}{c|}{0.926} & \multicolumn{1}{c|}{0.961} & 51.204   & \multicolumn{1}{c|}{38.752} & \multicolumn{1}{c|}{0.966} & \multicolumn{1}{c|}{0.981} & 31.756   \\ \hline
	\end{tabular}}\label{CVTC1}
\end{table*}
\subsection{Tensor robust principal component analysis}
In this section, we evaluate the performance of the proposed TRPCA method through HSI mixed noise denoising. The comparative TRPCA methods include the SNN \cite{2252014253}, TNN \cite{8606166}, 3DTNN and 3DLogTNN \cite{7342019749} methods.
\subsubsection{HSI denoising}
\begin{figure*}[!h] 
	\centering  
	\vspace{0cm} 
	\subfloat[]{
		\begin{minipage}[h]{0.13\linewidth}
			\includegraphics[width=1\linewidth]{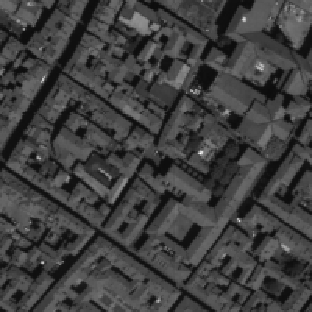}\\
			
			\includegraphics[width=1\linewidth]{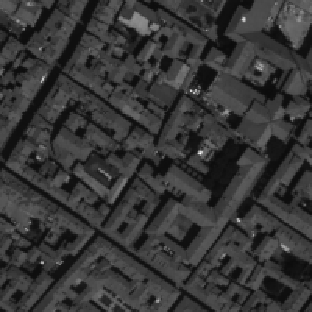}\\
			
			\includegraphics[width=1\linewidth]{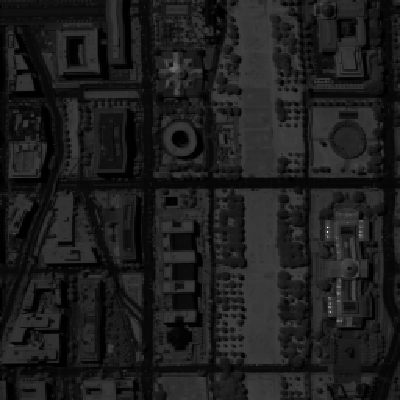}
		\end{minipage}
	}
	\subfloat[]{
	\begin{minipage}[h]{0.13\linewidth}
		\includegraphics[width=1\linewidth]{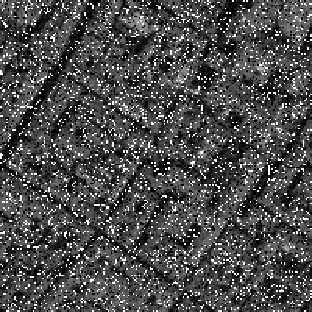}\\
		
		\includegraphics[width=1\linewidth]{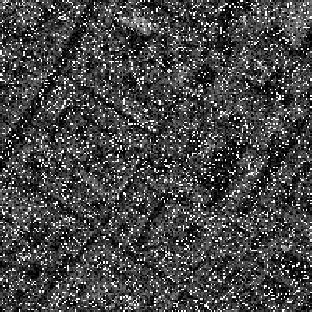}\\
		
		\includegraphics[width=1\linewidth]{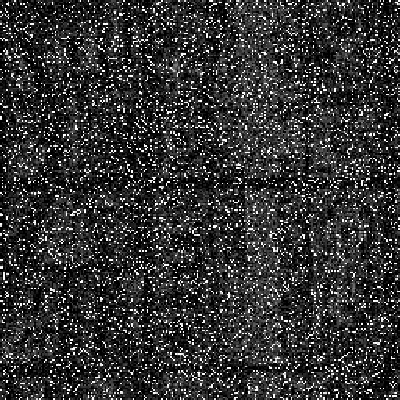}
	\end{minipage}
}
	\subfloat[]{
	\begin{minipage}[h]{0.13\linewidth}
		\includegraphics[width=1\linewidth]{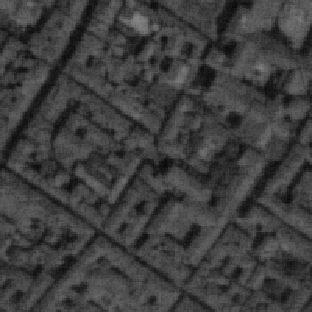}\\
		
		\includegraphics[width=1\linewidth]{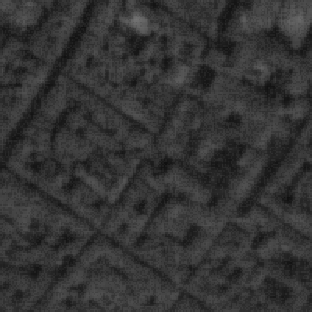}\\
		
		\includegraphics[width=1\linewidth]{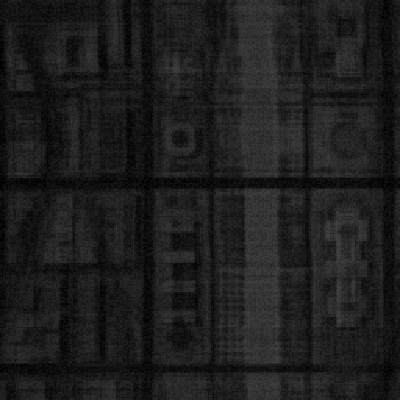}
	\end{minipage}
}
	\subfloat[]{
	\begin{minipage}[h]{0.13\linewidth}
		\includegraphics[width=1\linewidth]{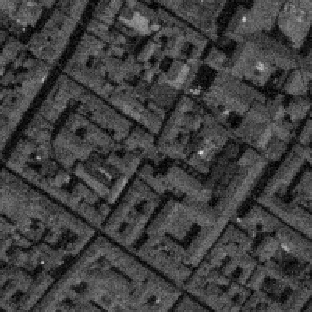}\\
		
		\includegraphics[width=1\linewidth]{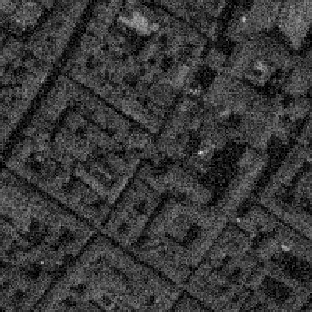}\\
		
		\includegraphics[width=1\linewidth]{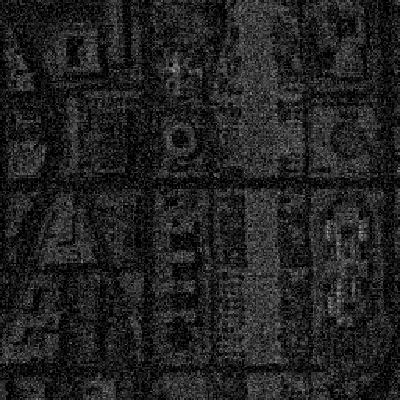}
	\end{minipage}
}
	\subfloat[]{
	\begin{minipage}[h]{0.13\linewidth}
		\includegraphics[width=1\linewidth]{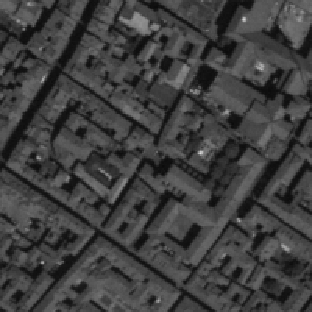}\\
		
		\includegraphics[width=1\linewidth]{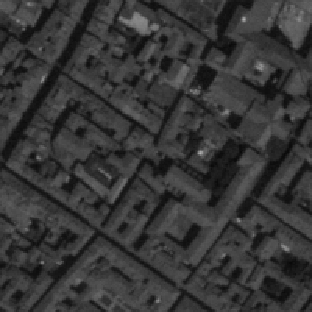}\\
		
		\includegraphics[width=1\linewidth]{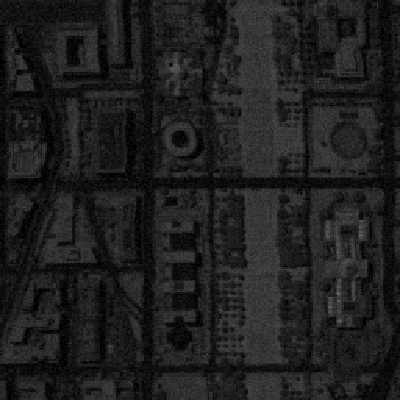}
	\end{minipage}
}
	\subfloat[]{
	\begin{minipage}[h]{0.13\linewidth}
		\includegraphics[width=1\linewidth]{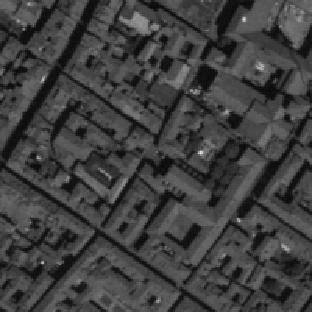}\\
		
		\includegraphics[width=1\linewidth]{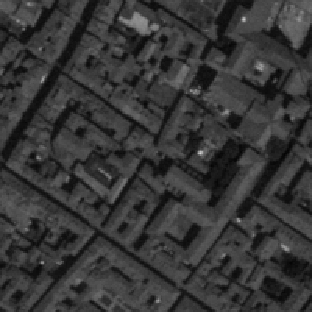}\\
		
		\includegraphics[width=1\linewidth]{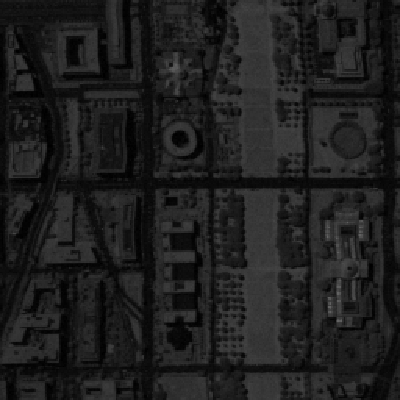}
	\end{minipage}
}
	\subfloat[]{
	\begin{minipage}[h]{0.13\linewidth}
		\includegraphics[width=1\linewidth]{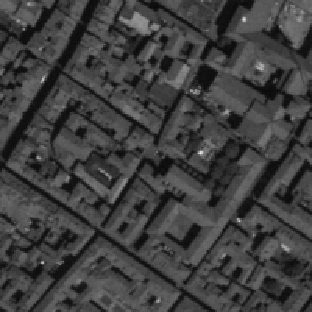}\\
		
		\includegraphics[width=1\linewidth]{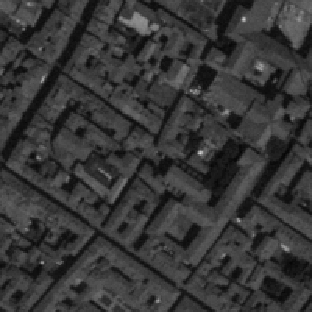}\\
		
		\includegraphics[width=1\linewidth]{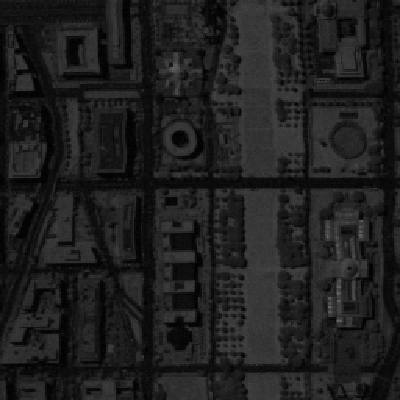}
	\end{minipage}
}
	\caption{(a) Original image. (b) Noise image. (c) SNN. (d) TNN. (e) 3DTNN. (f) 3DLogTNN. (g) EMLCP }
	\label{HSITRPCA}
\end{figure*}

We test the Pavia University data sets and Washington DC Mall data sets, where Pavia University data size is $200\times200\times80$ and Washington DC Mall data size is $256\times256\times150$. We divide the mixed noise into two kinds, one is independent identically distributed Gaussian noise plus independent identically distributed pepper and salt noise, and the other is non i.i.d. Gaussian noise plus i.i.d pepper and salt noise, where $\sigma$ is pepper and salt noise and $\nu$ is Gaussian noise. In Table \ref{TRPCA001}, we list the quantitative numerical results of Pavia University and Washington DC Mall Data under 3 combinations of these two kinds of noise respectively. It can be seen that under the influence of the weakest noise , the PSNR value of the obtained results is 0.6 dB higher than that of the suboptimal method 3DLogTNN. Even under the influence of the most severe noise, the PSNR value of the obtained results is still better than the suboptimal method 3DLogTNN. In Fig.\ref{HSITRPCA}, we show the visual results of the two kinds of data in turn according to the order of noise levels in Table \ref{TRPCA001}. The corresponding spectral bands are 50, 30, 100 respectively. It is easy to find from the figure that our method has better denoising effect than the comparative method.
\begin{table*}[]
	\caption{The PSNR, SSIM and FSIM values for 2 HSIs tested by observed and the nine utilized LRTC methods.}
	\resizebox{\textwidth}{!}{
		\begin{tabular}{|c|c|c|c|c|c|c|c|c|}
			\hline
			HSI                                & Mixed noise                   &      & Noise  & SNN    & TNN    & 3DTNN  & 3DLogTNN & EMLCP  \\ \hline
			\multirow{9}{*}{Pavia City Center} & \multirow{3}{*}{$\sigma$ = 0.05 $\nu$ = 0.2}     & PSNR & 11.646 & 26.899 & 27.920 & 35.833 & 37.231   & 37.825 \\ \cline{3-9} 
			&                               & SSIM & 0.119  & 0.788  & 0.780  & 0.969  & 0.974    & 0.976  \\ \cline{3-9} 
			&                               & FSIM & 0.532  & 0.868  & 0.890  & 0.979  & 0.982    & 0.984  \\ \cline{2-9} 
			& \multirow{3}{*}{$\sigma$ = 0.1 $\nu$ = 0.2}      & PSNR & 11.198 & 24.290 & 22.647 & 31.946 & 33.647   & 33.890 \\ \cline{3-9} 
			&                               & SSIM & 0.105  & 0.632  & 0.527  & 0.928  & 0.943    & 0.945  \\ \cline{3-9} 
			&                               & FSIM & 0.493  & 0.789  & 0.778  & 0.952  & 0.962    & 0.963  \\ \cline{2-9} 
			& \multirow{3}{*}{$\sigma$ follows U(0.1-0.15) $\nu$ = 0.2} & PSNR & 10.846 & 23.441 & 20.905 & 30.623 & 32.469   & 32.528 \\ \cline{3-9} 
			&                               & SSIM & 0.095  & 0.566  & 0.432  & 0.903  & 0.927    & 0.927  \\ \cline{3-9} 
			&                               & FSIM & 0.473  & 0.757  & 0.732  & 0.937  & 0.951    & 0.952  \\ \hline
			\multirow{9}{*}{Washington DC}     & \multirow{3}{*}{$\sigma$ = 0.05 $\nu$ = 0.2}     & PSNR & 11.279 & 27.737 & 28.002 & 36.428 & 38.767   & 39.448 \\ \cline{3-9} 
			&                               & SSIM & 0.116  & 0.794  & 0.750  & 0.959  & 0.979    & 0.980  \\ \cline{3-9} 
			&                               & FSIM & 0.517  & 0.883  & 0.882  & 0.978  & 0.986    & 0.988  \\ \cline{2-9} 
			& \multirow{3}{*}{$\sigma$ = 0.1 $\nu$ = 0.2}      & PSNR & 10.866 & 25.328 & 22.875 & 31.391 & 35.017   & 35.332 \\ \cline{3-9} 
			&                               & SSIM & 0.103  & 0.671  & 0.510  & 0.872  & 0.950    & 0.951  \\ \cline{3-9} 
			&                               & FSIM & 0.476  & 0.822  & 0.768  & 0.937  & 0.969    & 0.971  \\ \cline{2-9} 
			& \multirow{3}{*}{$\sigma$ follows U(0.1-0.15) $\nu$ = 0.2} & PSNR & 10.549 & 24.528 & 21.165 & 29.175 & 33.621   & 33.790 \\ \cline{3-9} 
			&                               & SSIM & 0.094  & 0.623  & 0.424  & 0.798  & 0.935    & 0.935  \\ \cline{3-9} 
			&                               & FSIM & 0.457  & 0.796  & 0.723  & 0.905  & 0.961    & 0.963  \\ \hline
	\end{tabular}}\label{TRPCA001}
\end{table*}

\section{Conclusion}
In this paper, we propose MLCP function, a new non-convex function, which finds that the Logarithmic function is the upper bound of the MLCP function. It is theoretically guaranteed that the MLCP function can achieve better results for the minimization problem. The proposed function is directly applied to the tensor recovery problem, its explicit solution cannot be obtained, which is very unfavorable to the solution of the algorithm. To this end, we further put forward the corresponding equivalence theorem to settle this problem. We apply the equivalent weighted tensor $L\gamma$-norm to the LRTC and TRPCA problems, giving their EMLCP-based models respectively. According to the Kurdyka-Łojasiwicz property, we prove that the solution sequence of the proposed algorithm has finite length and converges globally to a critical point. Extensive experiments show that our method can achieve good visual and numerical quantitative results. The obtained numerical quantitative results outperform the NWSTNN method using Logarithmic function, which is consistent with our theoretical analysis. In addition, it is worth studying whether the MLCP function can be extended in more applications.


%





\ifCLASSOPTIONcaptionsoff
  \newpage
\fi

\bibliography{bibtex/bare_jrnl_cs}


%




\end{document}